\documentclass[aps,jmp,superscriptaddress,showkeys,showpacs]{revtex4}

\usepackage{amssymb,hyperref}

\begin{document}
\def\wPm{{$\widehat P$-matrix}}
\def\wPms{{$\widehat P$-matrices}}
\def\Bcal{{\cal B}}
\def\Scal{{\cal S}}
\def\real{\mathbb{R}}
\def\complex{\mathbb{C}}

\title{Flat bases of invariant polynomials and \wPms\ of $E_7$ and $E_8$}

\author{Vittorino Talamini}
\email[]{vittorino.talamini@uniud.it} \affiliation{Department of
Physics, Universit\`a di Udine, via delle Scienze 208, 33100
Udine, Italy}

\date{\today}

\begin{abstract}
\noindent Let $G$ be a compact group of linear transformations of
an Euclidean space $V$. The $G$-invariant $C^\infty$ functions can
be expressed as $C^\infty$ functions of a finite basic set of
$G$-invariant homogeneous polynomials, sometimes called an
integrity basis. The mathematical description of the orbit space
$V/G$ depends on the integrity basis too: it is realized through
polynomial equations and inequalities expressing rank and positive
semi-definiteness conditions of the \wPm, a real symmetric matrix
determined by the integrity basis. The choice of the basic set of
$G$-invariant homogeneous polynomials forming an integrity basis
is not unique, so it is not unique the mathematical description of
the orbit space too. If $G$ is an irreducible finite reflection
group, Saito, Yano and Sekiguchi in 1980 characterized some
special basic sets of $G$-invariant homogeneous polynomials that
they called {\em flat}. They also found explicitly the flat basic
sets of invariant homogeneous polynomials of all the irreducible
finite reflection groups except of the two largest groups $E_7$
and $E_8$. In this paper the flat basic sets of invariant
homogeneous polynomials of $E_7$ and $E_8$ and the corresponding
\wPms\ are determined explicitly. Using the results here reported
one is able to determine easily the \wPms\ corresponding to any
other integrity basis of $E_7$ or $E_8$. From the \wPms\ one may
then write down the equations and inequalities defining the orbit
spaces of $E_7$ and $E_8$ relatively to a flat basis or to any
other integrity basis. The results here obtained may be employed
concretely to study analytically the symmetry breaking in all
theories where the symmetry group is one of the finite reflection
groups $E_7$ and $E_8$ or one of the Lie groups $E_7$ and $E_8$ in
their adjoint representations.

 \end{abstract}


\maketitle
\noindent Copyright (2010) American Institute of Physics. This
article may be downloaded for personal use only. Any other use
requires prior permission of the author and the American Institute
of Physics. The following article appeared in J. Math. Phys. {\bf
51}, 023520, (2010) and may be found at
\url{http://link.aip.org/link/?JMP/51/023520}.

\section{Introduction\label{Intro}}

Orbit spaces are the natural domain of functions that are
invariant  under transformations of a given transformation group,
because invariant functions are constant on the orbits of the
group. Orbit spaces are then the natural place where to study
physical systems with symmetry~\cite{as1983,MicZhi2001} and they
may be used to better understand some aspects of structural phase
transitions or of spontaneous symmetry breaking mechanisms (some
examples are in Refs.~\cite{gufetal,sar-tal1998} and
\cite{sar-val2005}).

Let's assume that the symmetry group $G$ is a compact group acting
linearly on a vector space $V$. Without loss of generality one may
assume that $V$ is a (real) Euclidean space and that $G$ is a
group of real orthogonal matrices acting on $V$ with the matrix
multiplication.

By Hilbert's and Schwarz's
theorems~\cite{springer,schw-top,schw-ihes}, all $G$-invariant
polynomial functions (and, more generally, all $G$-invariant
$C^\infty$ functions) can be written as polynomial (or $C^\infty$)
functions of a minimal finite basic set of $G$-invariant
homogeneous polynomials. This minimal finite basic set is
sometimes called an integrity basis of the $G$ action in $V$,
especially in the physical literature, and its elements are
usually called basic polynomials. A complete mathematical
description of the orbit space $V/G$ and its stratification is
also obtained through an integrity basis~\cite{as1983,schw-ihes}.
Integrity bases are then the basic tool to describe invariant
functions and to describe orbit spaces.

An integrity basis $\Bcal$ of the $G$ action in $V$ is not unique,
one has in fact an infinite number of possible choices of the set
of basic polynomials in $\Bcal$. This implies an infinite number
of possible mathematical descriptions of the orbit space $V/G$,
each one of them, $\Scal(\Bcal)$, corresponding to a given choice
$\Bcal$ of the integrity basis. These different mathematical
descriptions of the orbit space $V/G$ do not differ topologically and
are stratified in the same manner; a possible mathematical
description $\Scal(\Bcal')$ is in fact a diffeomorphic image of
$V/G$ and of another possible mathematical description
$\Scal(\Bcal)$. To simplify, let's call orbit space of the
$G$ action in $V$ any mathematical description $\Scal$ of the
orbit space $V/G$, obtained through a given choice of the
integrity basis of the $G$ action in $V$.

In Ref.~\cite{SYS1980} Saito, Yano and Sechiguchi, following the
earlier results by Saito~\cite{Saito1993} (Ref.~\cite{Saito1993}
was written 1n 1979), discovered that, if $G$ is an irreducible
finite reflection group, it is possible to choose the set of basic
polynomials forming an integrity basis in a unique way (unique
except for possible multiplications of the basic polynomials by
constant factors), if one requires that the basic polynomials
satisfy some supplementary conditions. They called these special
bases {\em flat}. In the same paper they also found explicitly a
flat basis for each one of the irreducible finite reflection
groups, except for the two largest groups $E_7$ and $E_8$. It is
worth to say that both in Ref.~\cite{SYS1980} and in
Ref.~\cite{Saito1993} it is claimed that Yano found a flat system
of basic invariant polynomials for the group $E_7$, but his work,
to my knowledge, is still unpublished (it should be Ref. 39 of
Ref.~\cite{Saito1993}).

Saito was lead to define~\cite{Saito1993} the flat bases of
invariant polynomials trying to understand the many relations that
were known to exist between the theory of singularities and the
theory of finite reflection groups (mainly due to the works of
Arnol'd (Ref. \cite{arn1972} is a key article of a series) and
Brieskorn~\cite{bries1971,slo1980}). A relevant fact was the
isomorphism between the base space of the universal unfolding of
an isolated hypersurface singularity (that is the parameter space
of a versal deformation of the singularity) and the orbit space of
the action of a finite reflection group $W$ on the
complexification $\complex\otimes V$ of the real vector space $V$
in which $W$ acts effectively ($W$ is here the monodromy group of
the singularity). On one side Saito proved that on the base space
of the universal unfolding of an isolated hypersurface singularity
it is possible to define a flat structure, obtained through the
period map associated to
a primitive form. This flat structure implies the existence on the base space of the universal unfolding 
of a system of coordinates with a flat metric, i.e. with an
everywhere vanishing curvature. The coordinates of the base space
are consequently said flat coordinates. The original references
are Ref. 27 of Ref.~\cite{Saito1993} (unpublished) and
Ref.~\cite{Saito1983} (all reviewed in Ref.~\cite{oda1987}). On
the other side he proved, just using standard properties of finite
reflection groups, that on the orbit space of the action of an
irreducible finite reflection group on the complexification of the
real vector space where it acts effectively, it is possible to
choose a system of coordinates (to be more precise, an integrity
basis for the ring of invariant polynomials that determines the
system of coordinates in the orbit space by identifying the basic
polynomials with the coordinates) with a real and constant flat
metric. It is said consequently that this system of coordinates
for the orbit space is a flat system of coordinates and that with
it the orbit space acquires a flat structure. It is also said that
the corresponding basic polynomials and the whole ring of
invariant polynomials acquire a flat structure and that the set of
basic polynomials is a flat set of basic polynomials. The original
references about this construction of the flat structure on the
set of invariant polynomials are Refs.~\cite{SYS1980,Saito1993}.

In 1991 Dubrovin, trying to give a coordinate independent
formulation of the Witten--Dijkstra--Verlinde--Verlinde
associativity equations that appeared those years in the setting
up of the two dimensional topological field theory, was lead to
define a geometric object that he called Frobenius manifold. He
also proved that the (complexified) orbit space of a finite
reflection group, endowed with the flat structure introduced by
Saito, satisfies all the axioms of a Frobenius
manifold~\cite{Dub1999,Dub1996} (in effect this is the most
elementary example of Frobenius manifold).

Recent articles describing the set up of the flat structure and
the Frobenius manifold structure on the (complexified) orbit space
of a finite reflection group and on the base space of the
universal unfolding of an isolated hypersurface singularity are
Refs.~\cite{Saito2004} and \cite{SaitoTak2008}. It is now widely
known that the theory of Frobenius manifolds is related to
different branches of mathematics and physics: singularity theory,
Coxeter groups theory, quantum cohomology, theory of integrable
systems, \ldots, and that these relations often allow to make a
substantial progress in the understanding of some aspects of one
or another of these theories. For this reason this is nowadays a
field of intensive theoretical research (see for example the
articles contained in the books of Ref.~\cite{Saito2004}
and~\cite{SaitoTak2008}).

The main goal of this paper is to complete the determination,
started in Ref.~\cite{SYS1980}, of the flat bases of invariant
polynomials of the irreducible finite reflection groups. In this
paper two flat bases of invariant polynomials for the groups $E_7$
and $E_8$ are explicitly determined. They have been obtained, as
suggested in Ref.~\cite{SYS1980}, through the calculation of the
\wPm, a real symmetric matrix, whose matrix elements are
polynomials of the basic polynomials. This paper also gives the
explicit expressions of the \wPms\ corresponding to the flat bases
of $E_7$ and $E_8$ that are quite difficult to calculate and
necessary both to obtain the flat bases and to describe
mathematically the orbit spaces.

The flat structures on the base space of the universal unfolding
of the isolated hypersurface singularities associated to the
reflection groups $E_7$ and $E_8$ have been obtained since a long
time by Yano (Ref. 38 of Ref.~\cite{Saito1993}, that seems to be
unpublished) and Kato and Watanabe~\cite{KatoWat1981}, following
Saito's theory of primitive forms~\cite{Saito1983,oda1987}.
However, these flat structures on the base space of the universal
unfolding of the isolated singularities of type $E_7$ and $E_8$ do
not allow to determine flat basic sets of invariant polynomials
for the groups $E_7$ and $E_8$. The flat structure on the rings of
invariant polynomials of $E_7$ and $E_8$ can be defined only with
the results of the present article. In fact, the \wPms\
corresponding to the flat bases of invariant polynomials of $E_7$
and $E_8$ here obtained give straightforwardly flat metric
matrices for the flat structures carried by the rings of invariant
polynomials and the corresponding (complexified) orbit spaces.

In the rest of this Introduction I wish to sketch what
could be a possible application of the results here obtained to
study spontaneous symmetry breakings  (i.e. structural
phase transitions) in physical theories whose symmetry group is
one of the finite reflection groups $E_7$ and $E_8$ or one of the
Lie groups $E_7$ and $E_8$ in their adjoint representations (the
finite reflection groups $E_7$ and $E_8$ are the Weyl groups of
the Lie groups $E_7$ and $E_8$).

The Lie group $E_8$ was fist proposed as the symmetry group of a
grand unification theory (GUT) in Ref.~\cite{barsgun1980}. After a
few years it appeared also in superstring theories, precisely in
the heterotic string theory based on the group $E_8\times E_8$
\cite{grossetal1985}. Since then many different variants of models
of GUT's and superstring theories based on the Lie group $E_8$
were proposed. In most of these models the basic fields of the
theory transform under the adjoint representation of $E_8$, which
has 248 dimensions. Spontaneous symmetry breaking mechanisms take
then place and reduce the true (residual) symmetry group from
$E_8$ to one of its subgroups in such a way to recover the
features of the standard model of elementary particles. I don't
want to enter here into the physics underlying these symmetry
breakings but I wish to remind how the results here obtained for
the finite reflection groups $E_7$ and $E_8$ can be used in the
study the symmetry breakings either of the finite reflection
groups $E_7$ and $E_8$, either of the Lie groups $E_7$ and $E_8$
in their adjoint representations.

It is well known that there is a one to one correspondence between
the bases of homogeneous invariant polynomials for the adjoint
representation Ad$(G)$ of a simple Lie group $G$ and the bases of
homogeneous invariant polynomials for the Weyl group $W$ of $G$,
that is a finite irreducible reflection group. The way to obtain a
basis ${\cal B}_W$ of $W$-invariant polynomials from a basis
${\cal B}_G$ of Ad$(G)$-invariant polynomials is to form ${\cal
B}_W$ with the polynomials obtained by the restrictions of the
basic polynomials in ${\cal B}_G$ to the Cartan subalgebra of the
Lie algebra of $G$ (details are given in Ref.~\cite{sar-tal1998},
Section III). In this way the bases ${\cal B}_G$ and ${\cal B}_W$
contain the same number $l$ of elements (where $l$ is the rank of
$G$ and of $W$) with the same degrees, but the basic polynomials
in ${\cal B}_W$ are defined in terms of only $l$ variables, while
those of ${\cal B}_G$ are defined in terms of dim(Ad$(G)$)
variables. An important fact is that the \wPm\ obtained from the
basis ${\cal B}_G$ of Ad$(G)$ is identical to the \wPm\ obtained
from the basis ${\cal B}_W$ of $W$ (Ref. \cite{sar-tal1998},
Section III), and this implies also the same geometric structure
(that is the same geometric stratification) of the orbit spaces of
the two group actions. Obviously, to a basis transformation ${\cal
B}_G\to{\cal B}'_G$ corresponds a basis transformation ${\cal
B}_W\to{\cal B}'_W$ obtained by the same transforming equations,
and in the two cases the orbit spaces transform in the same way.
All this implies that, for every choice of the basis ${\cal B}_G$,
there is a basis ${\cal B}_W$, such that the orbit spaces of the
actions of Ad$(G)$ and of $W$ are represented by the same
geometric subset $\Scal\subset \real^l$.

The orbit space $\Scal$ is a connected semi-algebraic subset of
$\real^l$ (where $l$ is the rank of the group), stratified, as all
semi-algebraic sets, in (connected) primary strata. Each one of
these primary strata of dimension greater than 0 is defined
through polynomial equations and inequalities, and its boundary is
formed by lower dimensional primary strata. This stratification of
$\Scal$ in primary strata coincides with the stratification due to
the group action (that is determined by the isotropy subgroup
classes), that is, there is a one to one correspondence between the
connected components of the strata defined by the isotropy
subgroup classes and the primary strata of $\Scal$. Moreover, if
the strata $\Sigma_1$ and $\Sigma_2$ (of the space $V$ where the
group acts) correspond to the isotropy subgroup classes $[H_1]$
and $[H_2]$, with $H_1\subset H_2$, then the primary
strata of $\Scal$ that form the image of $\Sigma_2$ lie in the
boundary of the primary strata  of $\Scal$ that form the image of
$\Sigma_1$ (that is, lower dimensional primary strata
correspond to greater isotropy subgroups). Because of this
correspondence between the stratification of $\Scal$ in primary
strata and the stratification of $V$ due to the group action, each primary strata of
$\Scal$ represents a possible pattern of symmetry breaking.

All Ad$(G)$-invariant (and all $W$-invariant) polynomial (or
$C^\infty$)-functions are obtained in a unique way in terms of the
basic polynomials in ${\cal B}_G$ (in ${\cal B}_W$), so if one
has an Ad$(G)$-invariant function $f$, and one writes it in terms
of the $l$ Ad$(G)$-basic polynomials in ${\cal B}_G$, one obtains
a function $\widehat f$ of $l$ variables, by considering the basic
polynomials as variables. The same function $\widehat f$
represents then also a $W$-invariant function, if one considers
the $l$ variables as the $l$ basic polynomials in ${\cal B}_W$.
$\widehat f$ is a function defined in a certain domain $D\subseteq
\real^l$, but only its restriction to $\Scal$ has the same range
of the original function $f$.

Let $f$ be an invariant potential function and $\widehat f$ its
representative defined in $\real^l$. $\widehat f$ depends in general on
the $l$ variables corresponding to the basic polynomials and to a certain number of
other parameters related to the physical problem one is dealing
with. Let the minimum point of $\widehat f$ be in a given point of
the orbit space $\Scal$, in a given stratum corresponding to a
given conjugacy class of isotropy subgroups (of Ad$(G)$ or of $W$,
depending on which one of the two groups one is dealing with). If for
some reason the parameters in $\widehat f$ change with continuity,
the minimum point of $\widehat f$, in general, changes its position with
continuity, and it may happen that it changes stratum, changing in
this way the conjugacy class of its isotropy subgroup (and
consequently the residual symmetry), realizing in this
way what is called a spontaneous symmetry breaking (or second
order structural phase transition).

If one studies in the orbit space the minimization of invariant
functions (and the related spontaneous symmetry breakings) there
is no difference between the problem where the symmetry group is
Ad$(G)$ and the problem where the symmetry group is $W$ (this is
an example of the so-called universality of the study of
spontaneous symmetry breaking in the orbit space). In both cases
one has in fact to study the minimization of the same function
$\widehat f$ of $l$ variables in the same semi-algebraic connected
subset $\Scal\subset\real^l$ representing the orbit space. In
considering the finite group one may have a great advantage
because the dimensions of the space where to study the
minimization are drastically reduced (for $E_8$, for example, 248
is reduced to 8).

What one has to do to study the spontaneous symmetry
breaking in the orbit space $\Scal$ of the $W$ action is summarized in the following points
(details and some examples are given in Refs. \cite{sar-tal1998,gufetal} and \cite{sar-val2005}).\\
1) Determine the equations and inequalities defining the various
strata of the orbit space $\Scal$.\\
2) Determine the conjugacy classes of isotropy subgroups
corresponding to the various strata
of the orbit space $\Scal$.\\
3) Express the invariant function $f$ in terms of the integrity
basis, obtaining in this way a function $\widehat f$ of $l$
variables.\\
4) Determine the minima of $\widehat f$ in the orbit space
$\Scal$, and, in particular, for which values of the parameters in
$\widehat f$ these minima fall into one or another stratum of $\Scal$.\\
If one is interested on the adjoint representation of a Lie group
$G$ one also has to:\\
5) determine the basis transformation relating the integrity basis
${\cal B}_W$, arising from the restriction to the Cartan
subalgebra of the integrity basis ${\cal B}_G$, and the basis chosen to express the \wPm.\\
Point 1) is solved using rank and positive semi-definiteness conditions of the \wPm. The \wPm\ depends on the
particular integrity basis chosen and it may be useful to
transform the basis, either to simplify the \wPm, either
to use known results in another basis. This transformation of the \wPm\ is easily obtained through the \wPm\ transformation
formula. Point 2) is solved by
determining the isotropy subgroups in convenient points $x\in V$
($V$ is the space where $W$ acts), each one of them having image
in each one of the primary strata of the orbit space $\Scal$.
(This image, as will be clarified later on, is a point of
$\real^l$ that has for coordinates the values taken in $x$ by the
basic polynomials). It may be useful to remind that the points $x$
whose image belongs to the boundary of $\Scal$ must necessarily
lie in the set of the reflecting hyperplanes of $W$. Point 4) is
solved for example with the method of Lagrange multipliers to find
conditional minima on the various primary strata of $\Scal$.

This study of minimization of invariant functions can be done
analytically, that is with full precision, because from the \wPms\
one may determine all the equations and inequalities defining the
various strata of the orbit space. These
calculations are not straightforward but do not present conceptual
difficulties. In practice, one has to do with systems of algebraic equations and inequalities that may be quite complicated to handle.

When a symmetry breaking takes place, the conjugacy class of the
isotropy subgroups changes from $[H_1]$ to $[H_2]$, where
$H_1\subset H_2$ or $H_2\subset H_1$. The most probable changes
are from a singular stratum (i.e. in the boundary of $\Scal$) to
the principal stratum, where the isotropy group is the smallest
possible, usually trivial (i.e. $H_2\equiv\{I\}$), and in the
opposite way the most probable changes are from a stratum to a
singular stratum in which the isotropy subgroups have rank
increased by 1: ${\rm rank}(H_2)={\rm rank}(H_1)+1$. For
particular variations of the parameters, there are however
possible other changes of the isotropy subgroups. To clear this
fact let's consider a three dimensional example arising from an
irreducible reflection group $W$ of rank 3 (that is, $W$ is $A_3$
or $B_3$ or $H_3$). (For $E_7$ and $E_8$ one has to generalize
this to 7 or 8 dimensions). The orbit space $\Scal$ is in this
case topologically equivalent to a pyramid having infinite height,
the vertex at the origin of $\real^3$, and 3 lateral faces joined
along 3 edges. (The true orbit space has curved surfaces and
edges, whose equations depend on the group.) The interior of this
pyramid corresponds to the principal stratum, in each of the
bidimensional faces the isotropy subgroups are of rank one, in
each of the one-dimensional edges the isotropy subgroups are of
rank two, and at the origin the isotropy subgroup is $W$, and this
is the only point where the symmetry is unbroken. A function
$\widehat f$ that for a given value of the parameters has the
minimum at the origin, when the parameters vary, may shift the
location of the minimum to other points of the orbit space,
located in the faces or in the edges, or in the principal stratum.
It is clear that only particular variations of the parameters may
realize the shift from the origin to the other singular strata and
that the most probable shift is toward the principal stratum. With
the high dimensional orbit spaces like those of $E_7$ or $E_8$
various possibilities of symmetry breaking may take place, with a
great possibility of changing of the ranks of the isotropy
subgroups.

Next section reviews some mathematical results. In particular, the
description of orbit spaces and their stratifications through the
\wPms, the definition of a flat basis of invariant polynomials,
and the \wPm\ transformation formula relating the \wPms\
corresponding to different integrity basis. Section~\ref{calc}
describes the calculations done to obtain the results here
reported. Sections \ref{E7} and \ref{E8} report a flat basis of
homogeneous invariant polynomials and the corresponding \wPm\
elements for each one of the two finite reflection groups $E_7$
and $E_8$, respectively. The flat basic polynomials are given in
terms of the integrity bases calculated by Mehta in
Ref.~\cite{Mehta1988}.

In Ref.~\cite{EPAPS} one may find files ready to be downloaded
that contain, in a computer readable format, all the results
reported in Sections \ref{E7} and \ref{E8} and some other related
formulas.

\section{Some mathematical results}

In this review section I shall only consider essential (i.e. with
no non-zero fixed points) linear actions of an irreducible finite
reflection group $W$. Many of the results here reported are
however true also for the case of a general linear compact group
$G$. The reader interested to a review concerning linear actions
of a general compact group $G$ may read Refs.~\cite{as1983} and
\cite{schw-ihes} or the very short Section II of
Ref.~\cite{gufetal} or Section 3 of Ref.~\cite{sar-val2005}. For
the unreferenced results the reader may refer to
Refs.~\cite{as1983} and \cite{schw-ihes} and the references
therein.

Let $W$ be an irreducible finite (real) reflection group, that is
an irreducible finite group generated by reflections. These groups
were classified by Coxeter~\cite{cox1934} in the following types:
$A_l$, $l\geq 1$, $B_l$, $l\geq 2$, $D_l$, $l\geq 4$, $E_6$,
$E_7$, $E_8$, $F_4$, $H_3$, $H_4$, $I_2(m)$, $m\geq 5$. The number
appearing as a subscript is the rank $l$ of the group and is equal
to the dimension of the real Euclidean space $V$ on which the
group action is essential. If it is given an orthonormal basis in
$V$, $W$ acts in $V$ as a group of real orthogonal matrices. The
limitations exclude group isomorphisms (like $I_2(6)\simeq G_2$).

 Every reflection $g\in W$ leaves fixed an
$(l-1)$-dimensional hyperplane $\pi_g$ of $V$, called the
reflecting hyperplane corresponding to $g$, and sends each $x\in
V$ to its symmetric with respect to $\pi_g$. The hyperplane
$\pi_g$ has equation $l_g(x)=0$, where $l_g(x)$ is a linear form
of (some of) the variables $x_1,\ldots,x_l,$ of the elements $x\in
V$.

Let $R$ be the algebra of polynomials of the $l$ variables
$x_1,\ldots,x_l$. The group $W$ also acts as a group of
automorphisms of $R$, the action being the following:
$(gp)(x)=p(g^{-1}x),\ \forall\, x\in V,\ g\in W$ and $p\in R$. The
$W$-invariant polynomials satisfy the condition $gp=p,\ \forall\,
g\in W$. They form a polynomial algebra, $R^W$, with $l$
algebraically independent generators, that is there exist $l$
$W$-invariant polynomials $p_1,\ldots, p_l,$ such that for every
$p\in R^W$, there exists a unique polynomial $\widehat p$ in $l$
indeterminates such that $p(x)=\widehat p(p_1(x),\ldots,
p_l(x)),\;\forall x\in V$. For the algebraic independence of the
generators, an equation like $\widehat p(p_1(x),\ldots, p_l(x))=0$
cannot be identically satisfied in $V$.

The basic polynomial generators $p_1,\ldots, p_l$ of $R^W$ can in
all generality be chosen homogeneous and Coxeter~\cite{cox1951}
found for each of the irreducible finite reflection groups, their
degrees $d_1,\ldots,d_l$. The product $\prod_{a=1}^l d_a$ is equal
to the order of the group and the sum $\sum_{a=1}^l d_a$ is equal
to $N+l$, where $N$ is the number of reflections in $W$. One may
order the basic polynomials in such a way that $d_1\leq
d_2\leq\ldots \leq d_l$. In the case of an irreducible finite
reflection group the inequalities among the $d_a$ are all strict,
except for $D_l$, with even $l$, in which there are two different
basic polynomials of degree $l$. Moreover, given the above
ordering, one has the identities: $d_a+d_{l-a+1}=d_l+2,\ \forall
a=1,\ldots,l$, and these imply $N=l d_l/2$. One may also put
$p_1(x)=<\!\!x,x\!\!>=||x||^2$, where $<\cdot,\cdot>$ is the
canonical scalar product in $V$ (a natural choice for real
orthogonal essential actions).

Explicit sets of basic invariant polynomials for all the
irreducible finite reflection groups have been obtained by many
authors. I just mention the two relevant
papers~\cite{cox1951,Mehta1988}.

The choice of a set of $l$ basic invariant homogeneous polynomials
$p_1,\ldots, p_l$ generating $R^W$ is not unique. Any other set
$q_1,\ldots,q_l$, of $l$ algebraically independent homogeneous
invariant polynomials of the same degrees $d_1,\ldots,d_l$, forms
a possible set of basic invariant polynomials. The $q_a(x)$ may
differ from the $p_a(x)$ either for a different choice of the
(orthonormal) coordinate system in $V$, either because the
$q_a(x)$ can be expressed as polynomials in the $p_a(x)$. This
freedom on the choice of the set of generators of $R^W$ is often
the origin of many difficulties. A set of basic invariant
homogeneous polynomials generating $R^W$ is sometimes called an
{\em integrity basis}, especially in the physical literature.

With an integrity basis $\Bcal=\{p_1(x),\ldots,p_l(x)\}$ one may
calculate an $l \times l$ real symmetric matrix $P(x)$ which has
its elements $P_{ab}(x)$ that are the scalar products of the
gradients of the basic invariants:
$$P_{ab}(x) =<\! \nabla p_a(x) \cdot \nabla p_b(x) \!>=\sum_{i=1}^l
\frac{\partial p_a(x)}{\partial x_i}\;\frac{\partial
p_b(x)}{\partial x_i}\qquad \forall\ a,b=1,\ldots,l.$$

The matrix $P(x)$ can also be written in the following way:
$$P(x)=j^T(x)\, j(x),$$
in which $j(x)$ is the jacobian matrix of the transformation $x\to
p$ (that is, $j_{ia}(x)={\partial p_a(x)}/{\partial x_i}$,
$\forall\, i=1,\ldots,l,\ a=1,\ldots,l)$, and the exponent $T$
means transposition. A classic result by Coxeter~\cite{cox1951}
claims that $\det j(x)$ is a homogeneous polynomial of degree $N$
of $x\in V$ proportional to the product of the linear forms
$l_g(x)$ whose vanishing determines the set of the reflecting
hyperplanes:
$$\det j(x)=c\, \prod_{g\in {\cal R}}\, l_g(x)\, ,$$
in which ${\cal R}$ is the set of reflections of $W$ and $c$ is a
constant. Then, $P(x)$ is a positive semi-definite matrix, in fact
$x^T P(x) x=||j(x) x||^2\geq 0\ \forall\ x\in V$, and one has
$\det P(x)=c^2\, \prod_{g\in {\cal R}}\, l_g^2(x)=0$ if $x$
belongs to the set of the reflecting hyperplanes and $\det P(x)>0$
otherwise.

From the homogeneity of the $p_a(x),\ \forall\,a=1,\ldots,l$, the
orthogonality of $W$, and the covariance of the gradients of the
$W$-invariant polynomials (that imply $\nabla' p(x')=g \nabla
p(x),\ x'=g x,\ \forall\,x\in V,\ g\in W\ \mbox{and}\ p\in R^W$),
the matrix elements $P_{ab}(x)$ are $W$-invariant homogeneous
polynomials of degree $d_a+d_b-2$, so that they may be expressed
as polynomials of the basic invariants: $$ P_{ab}(x) =\widehat
P_{ab}(p_1(x),\ldots,p_l(x)) \quad \forall x \in V.$$

A basic set ${\Bcal}$ of $l$ homogeneous invariant polynomials
$p_1(x),\ldots, p_l(x)$ determines a set of $l$ graded coordinates
$p_1,\ldots, p_l$ whose degrees are: $\deg(p_a)=\deg(p_a(x)),\
\forall a=1,\ldots,l$. Any invariant polynomial function of $x\in
V$, $p(x)$, determines a polynomial function of $p\in \real^l$,
$\widehat p(p)=\widehat p(p_1,\ldots,p_l)$, such that $\widehat
p(p_1(x),\ldots,p_l(x))=p(x),\ \forall x\in V$ (this actually
holds for any invariant $C^\infty$ function). If the invariant
polynomial $p(x)$ is homogeneous, then the polynomial $\widehat
p(p)$ is homogeneous, and $\deg(p(x))=\deg(\widehat p(p))$, taking
into account the degrees of the graded variables $p_a$.

As the matrix elements $P_{ab}(x)$ of the matrix $P(x)$ are
invariant homogeneous polynomial functions of $x\in V$, one can
define a matrix function of $p\in \real^l$, $\widehat P(p)$,
having $\widehat P_{ab} (p)=\widehat P_{ab} (p_1,\ldots,p_l)$ for
elements, such that $\widehat
P_{ab}(p_1(x),\ldots,p_l(x))=P_{ab}(x),\ \forall\, x\in V$. The
real symmetric matrix $\widehat P (p)$ is called the {\em
$\widehat P$-matrix} (associated to the given integrity basis
${\Bcal}$). The matrix elements of the \wPm\ are then homogeneous
polynomial functions of degree $d_a+d_b-2$ of the graded variables
$p_1,\ldots,p_l$. The first appearance of the \wPm\ has been in
Ref.~\cite{arn1976}, just for the groups $A_n$, and its first use
for an arbitrary reflection group has been made in
Refs.~\cite{arn1979,YaSek1979,Saito1993} and \cite{SYS1980}.
Sometimes the $\widehat P$-matrix is called the displacement
matrix or the discriminant matrix.

A change of the (orthonormal) coordinate system used in $V$: $x\to
x'=g x$, $g\in O(l)$, determines a change of the basic
polynomials: $p_a\to p_a'=g p_a$ (so $p_a'(x')=p_a(x)$), of the
jacobian matrix $j(x)$, of the equations of the reflecting
hyperplanes, and of the matrix elements of $P(x)$, viewed as
polynomials in $x$. This transformation of the basic polynomials,
consequent to a change of the (orthonormal) coordinate system in
$V$, doesn't determine changes in the matrix elements of the \wPm,
that is, using the basis $\{p\}$ or the basis $\{p'\}$, one
obtains the same expressions for the matrix elements of the \wPm:
$\widehat P_{ab}(p)=\widehat P'_{ab}(p')|_{p'\to p}$, where
$\widehat P_{ab}(p)$ and $\widehat P'_{ab}(p')$ are the \wPms\
calculated with the basic invariants $\{p'_a\}$ and $\{p_a\}$,
respectively. This follows because the equalities:
$p_a'(x')=p_a(x)$, if $x'=gx$, and the orthogonality of $g$,
imply: $P'_{ab}(x')=P_{ab}(x)$, if $x'=gx$, and because
$P_{ab}(x)=\widehat P_{ab}(p_1(x),\ldots,p_l(x))$ and
$P'_{ab}(x')=\widehat P'_{ab}(p'_1(x'),\ldots,p'_l(x'))=\widehat
P'_{ab}(p_1(x),\ldots,p_l(x))$, if $x'=gx$, imply $\widehat
P_{ab}(p)=\widehat P'_{ab}(p)=\widehat P'_{ab}(p')|_{p'\to p}$.
The \wPm\ is then dependent only on how a given integrity basis
$\{p_a\}$ generates $R^W$, and not on the coordinates $x$ used to
write explicitly the polynomials $p\in R^W$ (as $p_a(x),\
a=1,\ldots,l$). For this reason the \wPm\ seems to be a basic tool
of constructive invariant theory~\cite{tal-sspcm05}.

The map $\overline{p}: V\to \real^l:x\to
\overline{p}(x)=(p_1(x),\ldots,p_l(x))$ is called the {\em orbit
map} and maps $V$ onto a semi-algebraic connected proper subset
${\cal S}$ of $\real^l$. ${\cal S}$ can be identified with the
{\em orbit space} of the $W$ action in $V$ because there is a one
to one correspondence between the orbits in $V$ and the points in
${\cal S}$. This one to one correspondence is a consequence of the
fact that the basic polynomials separate the orbits, that is,
given any two orbits in $V$, at least one of the $p_a(x),\
a=1,\ldots,l$, takes a different value in the two given orbits.

Points lying in a same orbit of $V$ have conjugated isotropy
subgroups. The set of points in $V$ with conjugated isotropy
subgroups is called a {\em stratum} of the $W$ action in $V$. Its
image in the orbit space $V/W$ is called a {\em stratum} of the
orbit space $V/W$ and its image in $\Scal$ through the orbit map
is called a {\em stratum} of $\Scal$. Clearly there is a one to
one correspondence among strata of $V/W$ and strata of $\Scal$.
The number of strata is finite and equals the number of different
conjugacy classes of isotropy subgroups of the $W$ action in $V$.
As is outlined in the Introduction, the stratification (of $V$,
$V/W$ or $\Scal$) is relevant to study spontaneous symmetry
breaking mechanisms or structural phase transitions.

The \wPm\ contains all information needed to characterize
geometrically the connected set ${\cal S}\subset \real^l$. The
interior of ${\cal S}$ is the only subset of $\real^l$ where the
\wPm\ $\widehat P(p)$ is positive definite. The equation $\det
\widehat P(p)=0$ determines a closed surface
of $\real^l$ that contains the image, through the orbit map, of
the set of the reflection hyperplanes of $V$, and this image
coincides with the boundary of $\Scal$. The $k$-dimensional strata
of $\Scal$ are found from the equations and inequalities
expressing the conditions: $\mbox{rank}(\widehat P(p))=k$ and
$\widehat P(p)\geq 0$. These conditions may be obtained
conveniently, using the equations expressing the vanishing of all
minors of $\widehat P(p)$ of order $k+1$ and the inequalities
expressing the non negativity of all its principal minors of order
$\leq k$ (the non negativity of just the leading minors of a real
symmetric matrix is not a sufficient condition for its positive
semi-definiteness), taking care that at least one of the minors of
order $k$ be positive. $\Scal$ is a semi-algebraic set because all
these defining conditions are obtained through polynomial
equations and inequalities. One sees then that one has strata of
$\Scal$ of all dimensions $k$ such that $0\leq k\leq l$.
Simplifying the resulting system of algebraic
equations and inequalities, one may conveniently put $p_1=1$. In
fact, for the linearity of the action, one has $g(\lambda
x)=\lambda gx$, $\forall x\in V$ and $\lambda\in \real$, that
implies that the points $x$ and $\lambda x$, $\forall x\in V$ and
$\lambda\in \real,\ \lambda\neq 0$, have the same isotropy
subgroup and belong to the same stratum. This implies that the surface $||x||=1$ of $V$ intersects all strata of $V$
different from the origin, and that the hyperplane $p_1=1$ or $\real^l$
intersects all strata of $\Scal$ different from the origin. As a
consequence, the section of $\Scal$ in the hyperplane $p_1=1$ is a
compact connected subset of $\real^{l-1}$ that provides a concrete
representation of the orbit space $\Scal$. One may easily
reintroduce $p_1$ whenever one wishes because using all the $l$ variables
all equations and inequalities defining the strata of $\Scal$ must be homogeneous in the
graded variables $p_1,\ldots,p_l$.

The set $\Scal$ has dimension $l$ and, as all $l$-dimensional
semi-algebraic sets, has an $l$-dimensional interior, bordered by
connected $(l-1)$-dimensional faces (the primary strata), that are bordered by
$(l-2)$-dimensional faces, and so on, down to $0$-dimensional
vertices. All primary strata of $\Scal$ are contained in surfaces
determined by polynomial equations or by systems of polynomial
equations. Each $k$-dimensional primary stratum of $\Scal$, $\forall\,k>0$,
is open in its topological closure and its boundary is the union
of lower dimensional primary strata. A $k$-dimensional stratum of $\Scal$
is then the union of (some of the) $k$-dimensional primary strata of
$\Scal$. Moreover, if $x$ and $x'$ are points lying in different
strata of $V$, such that the isotropy subgroup $W_x$ is a proper
subgroup of the isotropy subgroup $W_{x'}$, then the stratum of
$\Scal$ containing $\overline{p}(x')$, where $\overline{p}$ is the
orbit map, lies in the boundary of the stratum of $\Scal$
containing $\overline{p}(x)$. Then, greater the isotropy
subgroups, smaller the dimension of the corresponding stratum of
$\Scal$. The origin of $V$ is the only stratum of $V$ with
isotropy subgroup $W$. Its image through the orbit map is the
origin of $\real^l$, that is the only 0-dimensional stratum of
$\Scal$, and belongs to the boundary of all the strata of $\Scal$.
A second order phase transition may take place only between
neighboring faces of $\Scal$, that is between neighboring strata
of $\Scal$, because of the continuity of the changing of the
location of the minimum point of a continuous invariant potential
function.

When one performs a change of the basic set of invariant
polynomials: $p_a\to q_a$ (in such a way to maintain the
homogeneity, the degrees $d_1,\ldots,d_l$, and the algebraic
independence of the basic polynomials), one determines a change of
the system of graded coordinates in $\real^l$, $p_a\to q_a$, that
implies a change of the \wPm\ and of the set $\Scal\subset
\real^l$ describing the orbit space. Here we are not considering
changes of the basic set of invariant polynomials $p_a\to p_a'=g
p_a$, $g\in O(l)$, corresponding to simple changes of the
orthonormal system of coordinates in $V$, because in that case the
\wPm\ and the set $\Scal$ would not change. Being the $q_a$
$W$-invariant polynomials, it is possible to write them as
polynomials of the $p_a$ (because the $p_a$ are forming an
integrity basis), in such a way that:
$$q_a(x)=\widehat q_a(p_1(x),\ldots,p_l(x)),\quad \forall\, x\in
V.$$

The jacobian matrix of this transformation, $J(p)$, has elements:
$$J_{ab}(p)=\frac{\partial \widehat q_b(p)}{\partial p_a},\qquad a,b=1,\ldots,l,$$
that are homogeneous polynomial functions of degree $d_b-d_a$ of
the graded variables $p_1,\ldots,p_l$, so they are vanishing if
$d_b<d_a$, and constant if $d_b=d_a$.

If all the degrees $d_a$ are different, that is, in the case of
any irreducible finite reflection group different from $D_l$, with
even $l$, this implies that $J(p)$ is an upper triangular matrix.
In the case of $D_l$, with even $l$, two degrees are equal and
there is a possible non-singular $2\times 2$ block along the
diagonal of $J(p)$. This would cause some small changes in all
what follows, but, as $D_l$ is not of our main interest in this
article, from now on we shall not consider $D_l$, with even $l$.

The determinant of $J(p)$ is then the product of the diagonal
elements $J_{aa}(p)$, that are all non zero constants, so $\det
J(p)\neq 0$, and the inverse transformation $q\to p$ is everywhere
well defined. The inverse transformation $q\to p$ is a polynomial
map, like the transformation $p\to q$, that is, the $\widehat
p_a(q),\ \forall a=1,\ldots,l$, are (homogeneous) polynomial
functions of the graded variables $q_1,\ldots,q_l$.

The transformation $p\to q$ causes a transformation of the \wPm\
and of the geometric shape of the orbit space ${\cal S}$. Because
of the regularity of the transformations $p\to q$ and $q\to p$,
this transformation of ${\cal S}$ is a diffeomorphism.

Let us write  $\widehat P_{ab}(q)$  and $\widehat P_{ab}(p)$,
$\forall\,a,b=1,\ldots,l$, the matrix elements of the \wPms\
$\widehat P(q)$ and $\widehat P(p)$ determined by the bases
$\{q\}$ and $\{p\}$, respectively. The relation between $\widehat
P(q)$ and $\widehat P(p)$ is obtained at once, in fact:

$$\widehat P_{cd}(q)|_{q\to \widehat q(p(x))}=<\!\nabla \widehat q_c(p(x)) \cdot \nabla
\widehat q_d(p(x))\!>=\sum_{a,b=1}^l \left<\!\left.\frac{\partial
\widehat q_c(p)}{\partial p_a}\right|_{p\to p(x)}\nabla p_a(x)
\cdot \left.\frac{\partial \widehat q_d(p)}{\partial
p_b}\right|_{p\to p(x)}\nabla p_b(x)\!\right>=$$
$$=\left.\sum_{a,b=1}^l J_{ac}(p)J_{bd}(p)\widehat
P_{ab}(p)\right|_{p\to p(x)}=\left.\sum_{a,b=1}^l
J^T_{ca}(p)\widehat P_{ab}(p)J_{bd}(p)\right|_{p\to p(x)}$$ that
is: $\widehat P(q)|_{q\to \widehat q(p)}=J^T(p)\cdot \widehat
P(p)\cdot J(p)$, that can better be written in the following way:
$$\widehat P(q)=\left.J^T(p)\cdot \widehat
P(p)\cdot J(p)\right|_{p\to \widehat p(q)}.$$ The formula just
obtained is called the {\em \wPm\ transformation formula}.

The transformation $p\to q$ implies a transformation of ${\cal S}$
and of all its strata. The equations and inequalities defining the
strata of $\Scal$ in the new variables $q$ may be obtained either
through the coordinate transformation $p\to q$ applied to the
equations and inequalities defining the strata of $\Scal$ in the
old variables $p$, either using the minors of the new \wPm\
$\widehat P(q)$. The minors of $\widehat P(q)$ themselves, in
general, are not obtained through the coordinate transformation
$p\to q$ applied to the minors of $\widehat P(p)$. The minors in fact
define surfaces of $\real^l$ that are different if calculated from
$\widehat P(p)$ or if calculated from $\widehat P(q)$, and these surfaces do not transform the ones into the others after the change of coordinates $p\to q$. Only
the intersections of these surfaces, that define the strata of
$\Scal$, define the same geometric objects in the two coordinate
systems.

The positiveness of the \wPm\ in the interior of ${\cal S}$ and
the just described transformation property allow to call the \wPm\
a metric matrix in the interior of ${\cal S}$. This metric is in
fact induced by the Euclidean metric of $V$.

The element $\widehat P_{ab}(p)$ of the \wPm\ has degree
(inherited from $P_{ab}(x)$):
$$\deg(\widehat P_{ab})=d_a+d_b-2=d_a+(d_l+2-d_{l-b+1})-2=d_a+d_l-d_{l-b+1}$$
so, the coordinate $p_l$ (of the maximum degree $d_l$)  is never
present in $\widehat P_{ab}(p)$ if $d_a<d_{l-b+1}$, that is, above
the anti-diagonal connecting $\widehat P_{1,l}$ and $\widehat
P_{l,1}$. Along this anti-diagonal $\deg(P_{ab})=d_l$ and under
this anti-diagonal $\deg(P_{ab})>d_l$.

In Refs.~\cite{SYS1980} and \cite{Saito1993} (and also in
Refs.~\cite{Dub1996} and \cite{Dub1999}) it is proved, in the case
of an irreducible finite reflection group $W$, that it is always
possible to choose a basis of $W$-invariant homogeneous
polynomials for which the matrix with elements
$$A_{ab}=\frac{\partial \widehat P_{ab}(p)}{{\partial p_l}},\qquad a,b=1,\ldots,l,$$
where $p_l$ is the basic invariant of the highest degree, is a
constant non-degenerate matrix. Moreover, given two bases for
which this property holds, their basic polynomials may differ only
for scale (i.e. multiplicative) constant factors. In
Refs.~\cite{SYS1980} and \cite{Saito1993} these bases were called
{\em flat}. For our previous discussion it follows that in a flat
basis the coordinate $p_l$ appears only in the anti-diagonal of
the \wPm\ (in fact, under this anti-diagonal ${\partial \widehat
P_{ab}(p)}/{{\partial p_l}}$ would have degree $d_a-d_{l-b+1}>0$
and would not be a constant). The constant non-degenerate matrix
$A$ is the flat metric matrix characterizing Saito's flat
structure~\cite{Saito1993} and the Frobenius manifold
structure~\cite{Dub1999} on the (complexified) orbit space.

To obtain the results presented in this article the matrix $A$ has been used just to find an integrity
basis in which it results a constant matrix. The integrity basis found
in this way is, by definition, a flat integrity basis.

\section{Description of the calculations \label{calc}}

In this section I describe the method to follow to find out a flat
basis of invariant polynomials for an irreducible reflection group
$W$.

One starts with a given basis $\{p_1(x),\ldots,p_l(x)\}$ of
invariant homogeneous polynomials for the group $W$. For the
groups $E_7$ and $E_8$ I used those suggested by Mehta in section
4 of Ref.~\cite{Mehta1988}, multiplied by some numeric factors.

One calculates then the matrix $P(x)$ and, expressing its matrix
elements in terms of the basic invariants, one finds out the
corresponding \wPm\ $\widehat P(p)$. The calculations to reach
these results for the groups $E_7$ and $E_8$ are very large. To
give an idea, the most complicated element in the matrix $P(x)$ of
$E_8$, $P_{8,8}(x)$, is an $E_8$-invariant homogeneous real
polynomial of degree $58$ in the $8$ variables $x_1,\ldots,x_8$.
This invariant polynomial must be equated to the most general
homogeneous real polynomial of degree $58$ that can be written in
function of the integrity basis of $E_8$ (of degree $58$ if
considered in the variables $x_1,\ldots,x_8$). As the elements of
an integrity basis of $E_8$ have degrees $2,8,12,14,18,20,24,30$,
this most general homogeneous polynomial of degree $58$ is: $z_1
p_8(x)p_7(x)p^2_1(x)+z_2 p_8(x)p_6(x)p_2(x)+\ldots +
z_{163}p_1^{29}(x)$, and depends linearly on $163$ arbitrary real
coefficients: $z_1,\ldots,z_{163}$. Expanding this algebraic
equation in terms of the variables $x_1,\ldots,x_8$ and collecting
the coefficients of similar monomials, one obtains a linear system
for the arbitrary coefficients $z_1,\ldots,z_{163}$, and solving
this system one gets the searched polynomial expansion:
$P_{8,8}(x)=\widehat P_{8,8}(p_1(x),\ldots,p_8(x))$ and the
polynomial expansion of the matrix element $\widehat P_{8,8}(p)$.

One writes out then the most general basis transformation $p\to q$
in such a way that $\widehat q_a(p)$ be homogeneous of degree
$d_a$, like $p_a$, of the graded variables $p_1,\ldots,p_l$. This
transformation contains several arbitrary coefficients
$z_1,z_2,\ldots$ that multiply the possible monomials in the
expressions. One may simplify a bit the transformation by fixing
the scale of the variables, that is by taking $\partial \widehat
q_a(p)/{\partial p_a}=1$, that is, $\widehat q_a(p)=p_a+\widehat
f_a(p_1,\ldots,p_{a-1})$, where $\widehat f_a(p_1,\ldots,p_{a-1})$
is the most general homogeneous polynomial function of degree
$d_a$ of the graded variables $p_1,\ldots,p_{a-1}$. For clarity,
for the group $E_8$, the most general transformation $p\to q$ of
this kind is: $\widehat q_1=p_1$, $\widehat q_2=p_2+z_1 p_1^4$,
$\widehat q_3=p_3+z_2 p_2 p_1^2+z_3 p_1^6$, and so on. (In the
case of $E_7$ one has 31 free parameters and in the case of $E_8$
one has 48 free parameters, if scale transformations are not
allowed).

One calculates then the  \wPm\ $\widehat P(q)$ from the \wPm\
transformation formula. $\widehat P(q)$ here depends on the free
parameters.

By requiring that $\partial \widehat P_{ab}(q)/{\partial q_l}=0$,
$\forall\ a,b=1,\ldots,l\ \mid a\neq l-b+1$, that is by requiring
that $q_l$ be present only along the anti-diagonal of the \wPm\
$\widehat P(q)$, one obtains a linear system for the free
parameters $z_1,z_2,\ldots$ that has a unique solution (unique
because of the constraints $\partial \widehat q_a(p)/{\partial
p_a}=1,\ \forall\,a=1,\ldots,l$, and the results in
Ref.~\cite{SYS1980}). This determines a transformation leading
from the basis $\{p\}$ to a flat basis $\{q\}$. This in turn gives
the corresponding \wPm\ $\widehat P(q)$, that I shall call the
{\em flat \wPm}.

One may now perform scale transformations $q_a\to c_a q_a$, with
the $c_a$ real coefficients, to obtain  for example integer (and
possibly small) coefficients in the matrix elements of the flat
\wPm\ $\widehat P(q)$.

To find the flat bases and the corresponding \wPms\ of the groups
$E_7$ and $E_8$ some shortcuts have been found, otherwise, the
requirements to have a reasonable computation time and to use a
reasonable amount of computer memory would not have been
satisfied, and it would not have been possible to reach the end of
the calculations.

Sections IV and V report the results of these calculations, that
is, a flat basis of invariant homogeneous polynomials for each one
of the groups $E_7$ and $E_8$. It will be reported the starting
basic polynomials $p_a(x)$, the transformation $p\to q$ leading to
a flat basis of invariant polynomials $q_a(x)$ and the
corresponding \wPm\ in the flat basis chosen. Due to the length of
these results the \wPm\ $\widehat P(p)$, corresponding to the
starting basic invariants $\{p_a(x)\}$, will not be written here,
but it is reported in the files in the supplementary material
repository~\cite{EPAPS}.

One can easily recover the original \wPm\ $\widehat P(p)$
from the inverse transformation $q\to p$ and the \wPm\
transformation formula $\widehat P(p)=\left.J^T(q)\cdot \widehat
P(q)\cdot J(q)\right|_{q\to \widehat q(p)}$, where
$$J(q)=J(p)^{-1}|_{p\to \widehat p(q)}$$ is the jacobian matrix of the
transformation $q\to p$. With the same kind of transformation one may obtain the \wPm\ in any other integrity basis.

The first row and column of the \wPm\ $\widehat P(p)$ (and
obviously of $\widehat P(q)$) are determined from Euler's theorem,
taking into account the homogeneity of the $p_a(x)$ and the
standard form of $p_1(x)$, and their elements are the following:
$$ \widehat P_{1,a}(p)= \widehat P_{a,1}(p)=2d_a p_a,\quad \forall\, a=1,\ldots,l.$$
Because of this fact and of the symmetry of the \wPm, in the
following sections only the matrix elements $\widehat P_{ab}(q)$
with $2\leq a\leq b\leq l,$ will be reported.

All relevant parts of the calculations here outlined are reported
in the files in the supplementary material repository~\cite{EPAPS}
ready to be downloaded and used in computer programs. In
particular the files report for each one of the two finite
reflection groups $E_7$ and $E_8$ the starting basic invariant
polynomials, the corresponding \wPm, the basis transformation to a
flat basis and its inverse, and the flat \wPm\ corresponding to
the flat basic invariants.

\section{Flat basic invariants and \wPm\ of $E_7$\label{E7}}
Let $f_r(x),\ r=1,\ldots,56$, the following linear forms of the
variables $x\in V$:
$$\pm x_i\pm x_j\pm x_k,\qquad \mbox{where:}$$
$$(i,j,k)=(1,2,7),\ (1,3,6),\ (1,4,5),\ (2,3,5),\ (2,4,6),\ (3,4,7),\
(5,6,7).$$ These forms are globally invariant under any
transformation of $E_7$. \\
A basis of algebraically independent invariant homogeneous
polynomials is the following:
$$p_a(x)=c_a\ \sum_{r=1}^{56} (f_r(x))^{d_a},\qquad a=1,\ldots,7$$
where
$c=\left(\frac{1}{24},\frac{1}{24},\frac{1}{8},\frac{1}{24},\frac{1}{24},\frac{1}{8},\frac{1}{24}\right)$
and $d=\left(2,6,8,10,12,14,18\right)$ are lists of coefficients
and exponents.\\ The numeric factors $c_a$ are not necessary and
are introduced just to obtain the $p_a(x)$ with integer
coefficients and a global numeric factor equal to 1. It follows,
for example, that $p_1(x)=\sum_{i=1}^7 x_i^2$.

A basis transformation $p\to q=\widehat q(p)$ to obtain a flat
basis of invariant polynomials is the following:\\

{\footnotesize\noindent
$\widehat q_1=p_1,$\\

\noindent$\widehat q_2=-
\left.\left(35 p_1^3-27p_2\right)\right/6,$\\

\noindent${\widehat q_3=
\left.\left(2555 p_1^4-3024 p_1 p_2+243
p_3\right)\right/60,}$\\

\noindent$\widehat q_4=-729 
\left.\left(7 p_1^5-14 p_1^2 p_2+3 p_1 p_3-2p_4\right)\right/70,$\\

\noindent$\widehat q_5=\left. \left( -62986 p_1^6-245322 p_1^3 p_2+ 248589 p_1^2 p_3-352836 p_1 p_4-15309 \left(11 p_2^2-4 p_5\right)\right) \right/ 630,$\\

\noindent$\widehat q_6=
\left.\left(311637755 p_1^7-491053563 p_1^4 p_2- 105671709 p_1^3
p_3+349852932 p_1^2 p_4+
132678 p_1 \left(2783 p_2^2-1032 p_5\right)-32805 \left(1001 p_2 p_3-216 p_6\right)\right)\right/200970,$\\

\noindent$\widehat q_7=
\left.\left(-1608536176703995 p_1^9+3765720816716166 p_1^6 p_2+
8366046538608 p_1^5 p_3-1414871708456250 p_1^4 p_4+320222727 p_1
\left(7767 p_3^2+\right.\right.\right.
\\ \left.\left. +33832 p_2 p_4\right) -1561518 p_1^3
\left(1584976833 p_2^2-563978620 p_5\right)+16061328
p_1^2  \left(20903047 p_2 p_3-4328100 p_6\right)+12557754 \left(4123945 p_2^3-\right.\right.\\
\left.\left.\left.-772497 p_3 p_4-2977380 p_2 p_5+510300
p_7\right)\right)\right/22229291700.$ \\
}

 Obviously, all bases that can be obtained from the basis
$\{q\}$ just written by multiplying the $q_a$ by arbitrary
constant factors, are all flat bases. I decided to choose these
normalizations to obtain integer coefficients in the \wPm\
$\widehat P(q)$ and the same coefficient (36) for the terms in
$q_7$ in the (anti-diagonal) matrix elements of $\widehat P(q)$.

The elements $\widehat P_{ab}(q)$, with $2\leq a\leq b\leq 7$, of
the \wPm\ in the flat basis chosen, are the following:\\

{\footnotesize
\noindent $\widehat P_{2,2}=4 q_1^5+20 q_1^2 q_2+20 q_1 q_3+10 q_4$\\

\noindent $\widehat P_{2,3}=6 \left(4 q_1^3 q_2+2 q_2^2-2 q_1^2 q_3+2 q_1 q_4+q_5\right)$\\

\noindent $\widehat P_{2,4}= 14 \left(2 q_1^4 q_2+2 q_1^3 q_3+4
q_2 q_3+2 q_1^2 q_4+q_1 \left(6
q_2^2+q_5\right)+q_6\right)$\\

\noindent $\widehat P_{2,5}= 16 \left(2 q_1^4 q_3+2 q_1^3 q_4+4
\left(q_3^2+q_2
q_4\right)+q_1^2 \left(6 q_2^2-q_5\right)+2 q_1 q_6\right)$\\

\noindent $\widehat P_{2,6}=36 \left(2 q_2^3+q_1^4 q_4+6 q_1 q_2
q_4+ 2 q_3 q_4+q_2 q_5+q_1^3 \left(2 q_2^2+q_5\right)+q_1^2
\left(6 q_2
q_3+q_6\right)+q_7\right)$\\

\noindent $\widehat P_{2,7}= 22 \left(6 q_2^2 q_4+4 q_1^3
\left(q_3^2+2 q_2 q_4\right)+q_4 q_5+q_1^2 \left(4 q_2^3+6 q_3
q_4+6 q_2 q_5\right) +2 q_1^4 q_6+2 q_3 q_6+2 q_1
\left(6 q_2^2 q_3+q_4^2-q_3 q_5+2 q_2 q_6\right)\right)$\\

\noindent $\widehat P_{3,3}= 4 q_1^7-14 q_1^4 q_2+28 q_1^3 q_3+14
q_1^2 q_4+
14 q_1 \left(4 q_2^2-q_5\right)+7 \left(-4 q_2 q_3+q_6\right)$\\

\noindent $\widehat P_{3,4}= 16 \left(2 q_1^5 q_2+2 q_1^4 q_3+4
q_2 q_4+q_1^2 \left(6
q_2^2+q_5\right)+q_1 \left(4 q_2 q_3+q_6\right)\right)$\\

\noindent $\widehat P_{3,5}= 18 \left(2 q_1^6 q_2-4 q_1^5 q_3+2
q_1^4 q_4+ q_1 \left(-8 q_3^2+4 q_2 q_4\right)+ 2 q_1^3 q_5+q_1^2
\left(24 q_2 q_3-q_6\right)+2
\left(2 q_2^3-q_2 q_5+q_7\right)\right)$\\

\noindent $\widehat P_{3,6}=20 \left(2 q_1^6 q_3+ 2 q_1^5 q_4+12
q_1^2 \left(q_3^2+q_2 q_4\right)+ q_1^4 \left(10
q_2^2-q_5\right)+4 q_1 \left(2 q_2^3+q_3 q_4+2 q_2 q_5\right)
+2 q_1^3 q_6+2 \left(6 q_2^2 q_3+q_4^2-q_3 q_5+q_2 q_6\right)\right)$\\

\noindent $\widehat P_{3,7}=4 \left(12 q_2^4+72 q_1^5 q_2 q_3-16
q_3^3+ 24 q_2 q_3 q_4-60 q_1^4 \left(q_3^2-q_2 q_4\right)+ 12
q_1^6 q_5+24 q_2^2 q_5-3 q_5^2+ 8 q_1^3 \left(10 q_2^3+6 q_3 q_4-3
q_2 q_5\right)+ 6 q_4 q_6+\right. \\ \left. +36 q_1^2 \left(q_3
q_5+q_2 q_6\right)+ 12 q_1 \left(12 q_2 q_3^2+6 q_2^2 q_4+ q_4
q_5-q_3
q_6\right)\right)$\\

\noindent $\widehat P_{4,4}=4 \left(2 q_1^9+18 q_1^6 q_2+108 q_1^3
q_2^2+ 18 q_1^4 q_4+36 q_1 \left(q_3^2+2 q_2 q_4\right)+ 9 q_1^2
\left(12 q_2 q_3+q_6\right)+ 3 \left(16
q_2^3+6 q_3 q_4+6 q_2 q_5+3 q_7\right)\right)$\\

\noindent $\widehat P_{4,5}= 40 \left(2 q_1^7 q_2+2 q_1^6 q_3+8
q_2^2 q_3+12 q_1^2 q_2 q_4+q_4^2+q_1^4 \left(10 q_2^2+q_5\right)+
2 q_1 \left(6 q_2^3+4 q_3 q_4+q_2 q_5\right)+2 q_2 q_6+q_1^3
\left(8 q_2 q_3+q_6\right)\right)$\\

\noindent $\widehat P_{4,6}= 44 \left(2 q_1^8 q_2+2 q_1^7 q_3+2
q_1^6 q_4+4 q_1^3 \left(q_3^2+6 q_2 q_4\right)+q_1^5 \left(18
q_2^2+q_5\right) +6 q_1^2 \left(6 q_2^3+2 q_3 q_4+q_2 q_5\right)+
q_1^4 \left(20 q_2 q_3+q_6\right)+ 2 q_1 \left(18 q_2^2 q_3+2
q_4^2+\right.\right.\\ \left.\left.+ q_3 q_5+ 3 q_2 q_6\right)+2
\left(4 q_2 q_3^2+8 q_2^2 q_4+q_4 q_5+q_3
q_6\right)\right)$\\

\noindent $\widehat P_{4,7}= 26 \left(16 q_1^7 q_2^2+28 q_1^6 q_2
q_3+24 q_2^3 q_3+ 4 q_1^8 q_4+8 q_3^2 q_4+ 12 q_1^5 \left(q_3^2+2
q_2 q_4\right)+10 q_1^4 q_2 \left(6 q_2^2+q_5\right)+4 q_2 \left(2
q_4^2+q_3 q_5\right)+ 6 q_2^2 q_6+q_5 q_6+\right.\\
\left.+8 q_1^3 \left(10 q_2^2 q_3+q_4^2+q_3 q_5+q_2 q_6\right)+ 6
q_1^2 \left(4 q_2 q_3^2+12 q_2^2 q_4+q_3 q_6\right)+ q_1 \left(36
q_2^4+48 q_2 q_3 q_4+12 q_2^2 q_5+q_5^2+4 q_4 q_6\right)\right)$\\

\noindent $\widehat P_{5,5}= 4 \left(4 q_1^{11}-22 q_1^8 q_2+44
q_1^7 q_3+22 q_1^6 q_4+88 q_1^3 \left(4 q_3^2+q_2 q_4\right)+22
q_1^5 \left(12 q_2^2-q_5\right)+44 q_1^2 \left(2 q_2^3+3 q_2
q_5\right)+
44 q_1 \left(12 q_2^2 q_3+q_4^2-2 q_3 q_5\right)+\right.\\
\left.+11 q_1^4 \left(-20 q_2 q_3+q_6\right)+ 44 \left(-4 q_2
q_3^2+2 q_2^2 q_4+q_3 q_6\right)\right)$\\

\noindent $\widehat P_{5,6}=8 \left(24 q_1^9 q_2+60 q_2^4-12 q_1^8
q_3+ 288 q_1^5 q_2 q_3-32 q_3^3+12 q_1^7 q_4+ 96 q_2 q_3 q_4-60
q_1^4 \left(q_3^2-2 q_2 q_4\right)+ 16 q_1^3 \left(25 q_2^3+3 q_3
q_4\right)+36 q_2^2 q_5- 3 q_5^2+\right.\\
\left. +6 q_1^6 \left(14 q_2^2+q_5\right)+ 12 q_1 \left(24 q_2
q_3^2+18 q_2^2 q_4+q_4 q_5\right)+ 12 q_4 q_6+36
q_1^2 \left(4 q_2^2 q_3+q_4^2+2 q_3 q_5+q_2 q_6\right)\right)$\\

\noindent $\widehat P_{5,7}= 28 \left(36 q_1^8 q_2^2+8 q_1^{10}
q_3-32 q_1^7 q_2 q_3+ 48 q_2^2 q_3^2+24 q_2^3 q_4+ 8 q_3 q_4^2+28
q_1^6 \left(q_3^2+q_2 q_4\right)+ 20 q_1^4 q_3 \left(12
q_2^2-q_5\right)-
8 q_3^2 q_5+4 q_2 q_4 q_5+\right.\\
\left.+4 q_1^5 \left(14 q_2^3+6 q_3 q_4+3 q_2 q_5\right)+ q_6^2+8
q_1^3 \left(-10 q_2 q_3^2+ 10 q_2^2 q_4+q_4 q_5+q_3 q_6\right)+
q_1^2 \left(100 q_2^4+64 q_3^3+48 q_2 q_3 q_4+3 q_5^2+6 q_4
q_6\right)+\right.\\ \left. +2 q_1 \left(16 q_2^3 q_3+12 q_2
\left(q_4^2+2 q_3 q_5\right)+6 q_2^2
q_6-q_5 q_6\right)\right)$\\

\noindent $\widehat P_{6,6}= 4 \left(4 q_1^{13}+52 q_1^{10}
q_2+728 q_1^7 q_2^2+ 52 q_1^9 q_3+364 q_1^6 q_2 q_3+ 26 q_1^8
q_4+156 q_1^5 \left(2 q_3^2+3 q_2 q_4\right)+
130 q_1^4 \left(14 q_2^3+2 q_3 q_4+q_2 q_5\right)+52 q_1^3 \left(50 q_2^2 q_3+\right.\right.\\
\left. \left.+3 q_4^2+q_2 q_6\right)+ 78 q_1^2 \left(4 q_2
q_3^2+18 q_2^2 q_4+ q_4 q_5+q_3 q_6\right)+ 26 q_1 \left(40
q_2^4+8 q_3^3+36 q_2 q_3 q_4+6 q_2^2 q_5+q_5^2+3 q_4 q_6\right)+
13
\left(40 q_2^3 q_3+8 q_3^2 q_4+ \right.\right. \\
\left.\left. +4 q_2  \left(4 q_4^2+ 3 q_3 q_5\right)+6 q_2^2
q_6+q_5 q_6\right)\right)$\\

\noindent $\widehat P_{6,7}= 20 \left(12 q_1^{12} q_2+60 q_1^9
q_2^2+108 q_1^8 q_2 q_3-24 q_1^7 \left(q_3^2-2 q_2 q_4\right)+ 14
q_1^6 \left(28 q_2^3+3 q_3 q_4\right)+ 18 q_1^5 \left(14 q_2^2
q_3+q_4^2+q_3 q_5\right)+15 q_1^4 \left(24 q_2 q_3^2+18 q_2^2
q_4+\right.\right.\\
\left. \left.+q_4 q_5\right)+ q_1^3 \left(420 q_2^4-40 q_3^3+240
q_2 q_3 q_4+ 60 q_2^2 q_5-3 q_5^2+12 q_4 q_6\right)+ 3 q_1^2
\left(200 q_2^3 q_3+12 q_3^2 q_4+36 q_2 q_4^2+6 q_2^2
q_6+3 q_5 q_6\right)+ 3 q_1 \left(24 q_2^2 q_3^2+\right. \right.\\
\left. \left.+72 q_2^3 q_4+12 q_3 q_4^2+12 q_3^2 q_5+q_6^2+ 12 q_2
\left(q_4 q_5+q_3 q_6\right)\right) + 2 \left(24 q_2^5+54 q_2^2
q_3 q_4+2 q_4^3+ 6 q_2^3 q_5+3 q_3 q_4 q_5+3 q_2 \left(8
q_3^3+q_5^2+3 q_4
q_6\right)\right)\right)$\\

\noindent $\widehat P_{7,7}=2 \left(8 q_1^{17}+816 q_1^{11} q_2^2+
2040 q_1^8 q_2^3+680 q_1^9 q_3^2+ 1904 q_1^6 q_2 \left(-q_3^2+q_2
q_4\right)+ 272 q_1^7 \left(18 q_2^2 q_3+q_4^2\right)+34 q_1^5
\left(196 q_2^4+28 q_3^3+84 q_2 q_3 q_4+\right.\right. \\
\left.\left. + 3 q_5^2\right)+340 q_1^4 \left(14 q_2^3 q_3+3 q_3^2
q_4+ 3 q_2 \left(q_4^2+q_3 q_5\right)\right)+ 68 q_1^3 \left(120
q_2^2 q_3^2+60 q_2^3 q_4-10 q_3^2 q_5+10 q_2 q_4 q_5+q_6^2\right)+
34
q_1^2 \left(84 q_2^5+120 q_2^2 q_3 q_4+4 q_4^3+ \right. \right.\\
\left. \left.+20 q_2^3 q_5+12 q_3 q_4 q_5+6 q_3^2 q_6+ q_2
\left(-40 q_3^3-3 q_5^2+12 q_4 q_6\right)\right)+ 34 q_1 \left(100
q_2^4 q_3+36 q_2^2 q_4^2+4
q_2^3  q_6+q_3 \left(16 q_3^3+3 q_5^2+6 q_4 q_6\right)+6 q_2 \left(4 q_3^2 q_4+\right.\right. \right.\\
\left.\left. \left.+q_5 q_6\right)\right)+ 17 \left(16 q_2^3
q_3^2+36 q_2^4 q_4+q_4 q_5^2+ 2 q_4^2 q_6-2 q_3 q_5 q_6+ 12 q_2^2
\left(q_4 q_5+q_3 q_6\right)+2 q_2
\left(12 q_3 q_4^2+12 q_3^2 q_5+q_6^2\right)\right)\right)$\\
}

\section{Flat basic invariants and \wPm\ of $E_8$\label{E8}}
Let $f_r(x),\ r=1,\ldots,240$, the following linear forms of the
variables $x\in V$:
$$\pm 2 x_i,\qquad \mbox{where:}\ i=1,2,\ldots,8;$$
$$\pm x_i\pm x_j\pm x_k\pm x_l,\qquad \mbox{where:}$$
$$(i,j,k,l)=(1,2,3,4),\ (1,2,5,6),\ (1,2,7,8),\ (1,3,5,7),\ (1,3,6,8),\ (1,4,6,7),\ (1,4,5,8),$$
$$ (2,3,5,8),\ (2,3,6,7),\ (2,4,5,7),\ (2,4,6,8),\
(3,4,5,6),\ (3,4,7,8),\ (5,6,7,8).$$ These forms are globally
invariant under any transformation of $E_8$. \\
A basis of algebraically independent invariant homogeneous
polynomials is the following:
$$p_a(x)=c_a\ \sum_{r=1}^{240} (f_r(x))^{d_a},\qquad a=1,\ldots,8$$
where
$c=\left(\frac{1}{120},\frac{1}{48},\frac{1}{48},\frac{1}{48},\frac{1}{48},\frac{1}{48},\frac{1}{48},\frac{1}{240}\right)$
and $d=\left(2,8,12,14,18,20,24,30\right)$ are lists of
coefficients and exponents.\\ The numeric factors $c_a$ are not
necessary and are introduced just to obtain the $p_a(x)$ with
integer coefficients and a global numeric factor equal to 1. It
follows, for example, that $p_1(x)=\sum_{i=1}^8 x_i^2$.

A basis transformation $p\to q=\widehat q(p)$ to obtain a flat
basis of invariant polynomials is the following:\\

{\footnotesize
\noindent $\widehat q_1=p_1,$\\

\noindent $\widehat
q_2=-15 
\left.\left(49 p_1^4-5 p_2\right)\right/16,$\\

\noindent $\widehat q_3=15
\left.\left(77077 p_1^6-9075 p_1^2 p_2+250
p_3\right)\right/56,$\\

\noindent $\widehat q_4=-75
\left.\left(1429571 p_1^7-176605 p_1^3 p_2+ 8450
p_1 p_3-900 p_4\right)\right/1232,$\\

\noindent $\widehat q_5=225
\left.\left( 19689050667 p_1^9-2366797290 p_1^5 p_2- 15497625 p_1
p_2^2+161551000 p_1^3 p_3- 26010000 p_1^2 p_4+330000
p_5\right)\right/128128,$\\

\noindent $\widehat q_6=-75
\left. \left(633116777839117 p_1^{10}- 74677421397150 p_1^6 p_2-
707886840375 p_1^2 p_2^2+5459892516000 p_1^4 p_3-
962711190000 p_1^3 p_4+\right.\right.\\
\left. \left.+19916370000 p_1 p_5+
1365000 \left(6137 p_2 p_3-1080 p_6\right)\right)\right/178610432,$\\

\noindent $\widehat q_7=1125 
\left.\left( 1337225890828995934221 p_1^{12}-
151828720025122620480 p_1^8 p_2-2256634263217351875 p_1^4
p_2^2+11690427987605112200 p_1^6 p_3-\right.\right. \\
\left.\left.- 2159165294739986400 p_1^5 p_4 - 7057177881390000 p_1
p_2 p_4 +58582842589110000 p_1^3 p_5 +1031550000 p_1^2
\left(68998291 p_2 p_3-6514452 p_6\right)-\right.\right.\\
\left. \left.
-56631250 \left(11608905 p_2^3+2090608 p_3^2-907200 p_7\right)\right)\right/9454564607488,$\\

\noindent $\widehat q_8=-1125 
\left.\left(462085649451433694986425168251 p_1^{15}
-52603136152311242821416393300 p_1^{11} p_2-765354001507546703732862525 p_1^7 p_2^2+\right.\right.\\
\left.\left. +4039554675366904849208253200 p_1^9
p_3-749025984522663001367215200 p_1^8
p_4-2159454965110111676898000 p_1^4 p_2
p_4+\right.\right.\\
\left.\left. +20857955298933465971874000 p_1^6 p_5 -34342066800000
p_1^2 \left(355763437 p_3 p_4+509733180 p_2 p_5\right)
+38965212000 p_1^5  \left(611209152292693 p_2 p_3-\right.\right.\right.\\
\left. \left. \left. -63983765280330 p_6\right) -143091945000 p_1
\left(119048012369 p_2^2 p_3-9892169712 p_4^2-21628086480 p_2
p_6\right) +139395750 p_1^3 \left(959882748247533 p_2^3- \right.\right.\right.\\
\left. \left. \left. -164000 \left(1227635657 p_3^2-947063502
p_7\right)\right) +1147961509200000
\left(688779 p_2^2 p_4+79895 p_3 p_5- 75600 p_8\right)\right)\right/1116546261885902848.$\\
}

 Obviously, all bases that can be obtained from the basis
$\{q\}$ just written by multiplying the $q_a$ by arbitrary
constant factors, are all flat bases. I decided to choose these
normalizations to obtain integer coefficients in the \wPm\
$\widehat P(q)$ and the same coefficient (60) for the terms in
$q_8$ in the (anti-diagonal) matrix elements of $\widehat P(q)$.

The elements $\widehat P_{ab}(q)$, with $2\leq a\leq b\leq 8$, of
the \wPm\ in the flat basis chosen, are the following:\\

{\footnotesize \noindent $\widehat P_{2,2}=15 q_1^7+70 q_1^3
q_2+7 q_1 q_3+7 q_4$\\

\noindent $\widehat P_{2,3}=9 \left(60 q_1^5
q_2+60 q_1 q_2^2+15 q_1^2 q_4+q_5\right)$\\

\noindent $\widehat P_{2,4}=10 \left(20 q_1^6 q_2+60 q_1^2 q_2^2+5
q_1^4 q_3+4 q_2 q_3+5 q_1^3 q_4+q_1
q_5+q_6\right)$\\

\noindent $\widehat P_{2,5}=12 \left(300 q_1^4 q_2^2+80 q_2^3+15
q_1^6 q_3+2 q_3^2+45 q_1^5 q_4+120 q_1 q_2 q_4+5 q_1^3 q_5+15
q_1^2 \left(4 q_2
q_3+q_6\right)+q_7\right)$\\

\noindent $\widehat P_{2,6}=13 \left(60 q_1^5 q_2^2+40 q_1^3 q_2
q_3+15 q_1^6 q_4+60 q_1^2 q_2 q_4+5 q_1^4 q_5+4 \left(q_3 q_4+q_2
q_5\right)+q_1 \left(80
q_2^3+q_7\right)\right)$\\

\noindent $\widehat P_{2,7}=30 \left(300 q_1^4 q_2 q_4+10 q_1^6
q_5+30 q_1^2 \left(q_3 q_4+2 q_2 q_5\right)+30 q_1^5 \left(2 q_2
q_3+q_6\right) +30 q_1 \left(4 q_2^2 q_3+q_4^2+2 q_2
q_6\right)+5 q_1^3 \left(80 q_2^3+2 q_3^2+q_7\right)+\right.\\
\left.+2
\left(60 q_2^2 q_4+q_3 q_5+q_8\right)\right)$\\

\noindent $\widehat P_{2,8}= 9 \left(160 q_2^3 q_3+120 q_2
q_4^2+q_5^2+60 q_1^5 \left(3 q_3 q_4+4 q_2 q_5\right)+40 q_1^3
\left(60 q_2^2 q_4+q_3 q_5\right)+120 q_2^2 q_6+75 q_1^4 \left(8
q_2^2 q_3+3 q_4^2+8 q_2 q_6\right)+\right.\\ \left.+60 q_1 \left(4
q_2 q_3 q_4+4 q_2^2 q_5+q_4 q_6\right)+30 q_1^6 q_7+2 q_3 q_7+30
q_1^2 \left(40 q_2^4+q_4 q_5+2
q_2 \left(2 q_3^2+q_7\right)\right)\right)$\\

\noindent $\widehat P_{3,3}=30 \left(30 q_1^{11}+440 q_1^3
q_2^2+33 q_1^5 q_3+44 q_2 q_4+11 q_1
q_6\right)$\\

\noindent $\widehat P_{3,4}=12 \left(300 q_1^8 q_2+600 q_1^4
q_2^2+160 q_2^3+120 q_1^2 q_2 q_3+30 q_1^5 q_4+120
q_1 q_2 q_4+10 q_1^3 q_5+q_7\right)$\\

\noindent $\widehat P_{3,5}=210 \left(60 q_1^{10} q_2+280 q_1^6
q_2^2+80 q_1^4 q_2 q_3+30 q_1^7 q_4+80 q_1^3 q_2 q_4+2 q_1^5 q_5+8
q_1 \left(q_3 q_4+q_2 q_5\right)+4 \left(4 q_2^2 q_3+q_4^2+2 q_2
q_6\right)+q_1^2
\left(320 q_2^3+q_7\right)\right)$\\

\noindent $\widehat P_{3,6}= 30 \left(1200 q_1^7 q_2^2+75 q_1^9
q_3+30 q_1^3 \left(40 q_2^3+q_3^2\right)+600 q_1^4 q_2 q_4+30 q_1
\left(8 q_2^2 q_3+q_4^2\right)+60 q_1^2 q_2 q_5+45 q_1^5
q_6+2 \left(60 q_2^2 q_4+q_8\right)\right)$\\

\noindent $\widehat P_{3,7}=1020 \left(100 q_1^9 q_2^2+80 q_1^7
q_2 q_3+15 q_1^{10} q_4+140 q_1^6 q_2 q_4+8 q_2 q_3 q_4+5 q_1^8
q_5+8 q_2^2 q_5+20 q_1^4 \left(q_3 q_4+q_2 q_5\right)+2 q_1^2
\left(120 q_2^2 q_4+q_3 q_5\right)+2 q_4 q_6+\right.\\ \left.+10
q_1^3 \left(12 q_2^2 q_3+q_4^2+4 q_2 q_6\right)+q_1^5 \left(560
q_2^3+q_7\right)+2 q_1 \left(80 q_2^4+q_4
q_5+q_2 \left(4 q_3^2+q_7\right)\right)\right)$\\

\noindent $\widehat P_{3,8}= 300 \left(128 q_2^5+15 q_1^8 \left(80
q_2^3+3 q_3^2\right)+600 q_1^9 q_2 q_4+1680 q_1^5 q_2^2 q_4+4 q_3
q_4^2+35 q_1^6 \left(32 q_2^2 q_3+3 q_4^2\right)+160 q_1^7 q_2
q_5+8 q_2 q_4 q_5+\right.\\
\left. +160 q_1^3 q_2 \left(2 q_3 q_4+q_2 q_5\right)+60 q_1^{10}
q_6+q_6^2+q_1^2 \left(480 q_2^3 q_3+4 q_3^3+120 q_2
q_4^2+q_5^2+240 q_2^2 q_6\right)+4 q_2^2 \left(4
q_3^2+q_7\right)+10 q_1^4 \left(280 q_2^4+q_4 q_5+3 q_3 q_6+
\right.\right.\\ \left.\left.+2 q_2 q_7\right)+2 q_1 \left(320
q_2^3 q_4+8 q_2 q_3
q_5+q_4 q_7\right)\right)$\\

\noindent $\widehat P_{4,4}=300 q_1^{13}+1300 q_1^9 q_2+9360 q_1^5
q_2^2+130 q_1^7 q_3+520 q_1^6 q_4+ 1560 q_1^2 q_2 q_4+52 \left(q_3
q_4+2 q_2 q_5\right)+130 q_1^3 \left(8 q_2 q_3+q_6\right)+13 q_1
\left(320 q_2^3+4
q_3^2+q_7\right)$\\

\noindent $\widehat P_{4,5}=15 \left(900 q_1^{11} q_2+4800 q_1^7
q_2^2+150 q_1^9 q_3+1800 q_1^4 q_2 q_4+40 q_1^6 q_5+120 q_1^2
\left(q_3 q_4+q_2 q_5\right)+30 q_1^5 \left(20 q_2
q_3+q_6\right)+60 q_1 \left(12 q_2^2 q_3+q_4^2+\right.\right.\\
\left.\left.+2 q_2 q_6\right)+5 q_1^3 \left(1280 q_2^3+8
q_3^2+q_7\right)+4 \left(120 q_2^2 q_4+q_3
q_5+q_8\right)\right)$\\

\noindent $\widehat P_{4,6}=16 \left(300 q_1^{12} q_2+1800 q_1^8
q_2^2+560 q_1^6 q_2 q_3+150 q_1^9 q_4+600 q_1^5 q_2 q_4+10 q_1^7
q_5+80 q_1^3 \left(q_3 q_4+q_2 q_5\right)+8 q_1 \left(90 q_2^2
q_4+q_3 q_5\right)+60 q_1^2 \left(8 q_2^2 q_3+\right.\right.\\
\left.\left.+q_4^2+2 q_2 q_6\right)+5 q_1^4 \left(800
q_2^3+q_7\right)+4 \left(80 q_2^4+q_4 q_5+q_2
\left(4 q_3^2+q_7\right)\right)\right)$\\

\noindent $\widehat P_{4,7}=24 \left(9900 q_1^{10} q_2^2+225
q_1^{12} q_3+960 q_2^3 q_3+4 q_3^3+675 q_1^{11} q_4+7200 q_1^7 q_2
q_4+360 q_2 q_4^2+75 q_1^9 q_5+3 q_5^2+ 90 q_1^5 \left(5 q_3
q_4+12 q_2 q_5\right)+\right.\\ \left.+60 q_1^3 \left(240 q_2^2
q_4+q_3 q_5\right)+360 q_2^2 q_6+225 q_1^8 \left(12 q_2
q_3+q_6\right)+225 q_1^4 \left(40 q_2^2 q_3+3 q_4^2+4 q_2
q_6\right)+ 90 q_1 \left(12 q_2 q_3 q_4+8 q_2^2 q_5+q_4
q_6\right)+3 q_3 q_7+\right.\\ \left.+15 q_1^6 \left(2240 q_2^3+14
q_3^2+q_7\right)+90 q_1^2 \left(200 q_2^4+q_4 q_5+q_3 q_6+q_2
\left(4
q_3^2+q_7\right)\right)\right)$\\

\noindent $\widehat P_{4,8}=21 \left(44000 q_1^9 q_2^3+3600
q_1^{11} q_2 q_3+9900 q_1^{10} q_2 q_4+720 q_2^2 q_3 q_4+ 20
q_4^3+300 q_1^{12} q_5+320 q_2^3 q_5+4 q_3^2 q_5+450 q_1^8 \left(3
q_3 q_4+2 q_2 q_5\right)+\right.\\ \left.+70 q_1^6 \left(480 q_2^2
q_4+q_3 q_5\right)+ 120 q_2 q_4 q_6+2400 q_1^7 q_2 \left(6 q_2
q_3+q_6\right)+150 q_1^4 \left(20 q_2 q_3 q_4+24 q_2^2 q_5+q_4
q_6\right)+q_5 q_7+ 20 q_1^3 \left(800 q_2^3 q_3+180 q_2
q_4^2+\right.\right.\\ \left.\left. +120 q_2^2 q_6+q_3
q_7\right)+15 q_1^2 \left(1280 q_2^3 q_4+8 q_3^2 q_4+16 q_2 q_3
q_5+2 q_5 q_6+q_4 q_7\right)+ 120 q_1 \left(80 q_2^5+q_3 q_4^2+2
q_2 \left(q_4 q_5+q_3 q_6\right)+q_2^2 \left(4
q_3^2+q_7\right)\right)+\right.\\ \left.+120 q_1^5 \left(560
q_2^4+2 q_4 q_5+q_2 \left(14 q_3^2+q_7\right)\right)\right)$\\

\noindent $\widehat P_{5,5}=15 \left(900 q_1^{17}+5100 q_1^{13}
q_2+61200 q_1^9 q_2^2+510 q_1^{11} q_3+2040 q_1^{10} q_4+14280
q_1^6 q_2 q_4+ 1020 q_1^4 \left(q_3 q_4+2 q_2 q_5\right)+136 q_1^2
\left(180 q_2^2 q_4+q_3 q_5\right)+\right.\\ \left. +510 q_1^7
\left(16 q_2 q_3+q_6\right)+680 q_1^3 \left(28 q_2^2 q_3+3 q_4^2+2
q_2 q_6\right)+136 \left(8 q_2 q_3 q_4+4 q_2^2 q_5+q_4
q_6\right)+51 q_1^5 \left(2240 q_2^3+12 q_3^2+q_7\right)+136 q_1
\left(200 q_2^4+\right.\right.\\ \left.\left. +q_4 q_5+q_3 q_6+q_2
\left(4
q_3^2+q_7\right)\right)\right)$\\

\noindent $\widehat P_{5,6}=18 \left(4500 q_1^{14} q_2+19800
q_1^{10} q_2^2+5400 q_1^8 q_2 q_3+450 q_1^{11} q_4+7200 q_1^7 q_2
q_4+150 q_1^9 q_5+120 q_1^5 \left(9 q_3 q_4+5 q_2 q_5\right)+ 80
q_1^3 \left(210 q_2^2 q_4+q_3 q_5\right)+\right.\\  \left.+ 150
q_1^4 \left(64 q_2^2 q_3+3 q_4^2+8 q_2 q_6\right)+120 q_1 \left(8
q_2 q_3 q_4+6 q_2^2 q_5+q_4 q_6\right)+ 15 q_1^6 \left(4480
q_2^3+q_7\right)+2 \left(640 q_2^3 q_3+240 q_2 q_4^2+q_5^2+120
q_2^2 q_6+2 q_3 q_7\right)+\right.\\ \left.+60 q_1^2 \left(440
q_2^4+2 q_4 q_5+q_2 \left(12
q_3^2+q_7\right)\right)\right)$\\

\noindent $\widehat P_{5,7}=1200 \left(300 q_1^{16} q_2+3900
q_1^{12} q_2^2+512 q_2^5+75 q_1^{14} q_3+45 q_1^8 \left(320
q_2^3+q_3^2\right)+ 75 q_1^{13} q_4+1800 q_1^9 q_2 q_4+8 q_3
q_4^2+15 q_1^{11} q_5+40 q_1^7 \left(3 q_3 q_4+\right.\right.\\
\left.\left.+4 q_2 q_5\right)+10 q_1^5 \left(504 q_2^2 q_4+q_3
q_5\right)+ q_6^2+15 q_1^{10} \left(44 q_2 q_3+q_6\right)+35 q_1^6
\left(96 q_2^2 q_3+3 q_4^2+4 q_2 q_6\right)+8 q_2 \left(2 q_4
q_5+q_3 q_6\right)+ 20 q_1^3 \left(28 q_2 q_3 q_4+\right.\right.\\
\left.\left.+16 q_2^2 q_5+q_4 q_6\right)+4 q_2^2 \left(8
q_3^2+q_7\right)+q_1^2 \left(1760 q_2^3 q_3+4 q_3^3+360 q_2
q_4^2+q_5^2+240 q_2^2 q_6+q_3 q_7\right)+2 q_1 \left(800 q_2^3
q_4+4 q_3^2 q_4+12 q_2 q_3 q_5+q_5 q_6+\right.\right.\\
\left.\left.+q_4 q_7\right)+10 q_1^4 \left(1400 q_2^4+3
q_4 q_5+2 q_3 q_6+q_2 \left(16 q_3^2+q_7\right)\right)\right)$\\

\noindent $\widehat P_{5,8}= 345 \left(9600 q_1^{15} q_2^2+62400
q_1^{11} q_2^3+4200 q_1^{13} q_2 q_3+1440 q_1^7 q_2 \left(80
q_2^3+q_3^2\right)+900 q_1^{16} q_4+3900 q_1^{12} q_2 q_4+330
q_1^{10} \left(q_3 q_4+2 q_2 q_5\right)+\right.\\
\left.+90 q_1^8 \left(360 q_2^2 q_4+q_3 q_5\right)+600 q_1^9
\left(22 q_2^2 q_3+q_4^2+q_2 q_6\right)+70 q_1^6 \left(48 q_2 q_3
q_4+32 q_2^2 q_5+3 q_4 q_6\right)+ 2 q_1^5 \left(13440 q_2^3
q_3+1260 q_2 q_4^2+4 q_5^2+840 q_2^2 q_6+\right.\right.\\
\left.\left.+3 q_3 q_7\right)+5 q_1^4 \left(6720 q_2^3 q_4+36
q_3^2 q_4+40 q_2 q_3 q_5+2 q_5 q_6+3 q_4 q_7\right)+ q_1^2
\left(3360 q_2^2 q_3 q_4+120 q_4^3+1280 q_2^3 q_5+240 q_2 q_4
q_6+q_5 \left(8 q_3^2+q_7\right)\right)+\right.\\ \left.+40 q_1^3
\left(1120 q_2^5+3 q_3 q_4^2+4 q_2 \left(3 q_4 q_5+2 q_3
q_6\right)+2 q_2^2 \left(16 q_3^2+q_7\right)\right)+ 2 q_1
\left(1760 q_2^4 q_3+720 q_2^2 q_4^2+8 q_3 q_4 q_5+320 q_2^3
q_6+q_6 q_7+4 q_2 \left(4 q_3^3+q_5^2+\right.\right.\right.\\
\left.\left.\left.+q_3 q_7\right)\right)+ 4 \left(400 q_2^4 q_4+12
q_2^2 q_3 q_5+q_4 \left(q_4 q_5+2 q_3 q_6\right)+2 q_2 \left(4
q_3^2 q_4+q_5 q_6+q_4
q_7\right)\right)\right)$\\

\noindent $\widehat P_{6,6}=2 \left(2250 q_1^{19}+171000 q_1^{11}
q_2^2+4275 q_1^{13} q_3+1140 q_1^7 \left(280 q_2^3+3
q_3^2\right)+51300 q_1^8 q_2 q_4+91200 q_1^4 q_2^2 q_4+1710 q_1^5
\left(56 q_2^2 q_3+3 q_4^2\right)+ \right.\\ \left. +5320 q_1^6
q_2 q_5+4560 q_1^2 q_2 \left(3 q_3 q_4+q_2 q_5\right)+1425 q_1^9
q_6+38 q_1 \left(480 q_2^3 q_3+4 q_3^3+120 q_2 q_4^2+q_5^2+120
q_2^2 q_6\right)+ 380
q_1^3 \left(680 q_2^4+2 q_4 q_5+3 q_3 q_6+\right.\right.\\
\left.\left.+q_2 q_7\right)+38 \left(320
q_2^3 q_4+8 q_2 q_3 q_5+q_4 q_7\right)\right)$\\

\noindent $\widehat P_{6,7}= 84 \left(4500 q_1^{17} q_2+31500
q_1^{13} q_2^2+154000 q_1^9 q_2^3+9000 q_1^{11} q_2 q_3+1125
q_1^{14} q_4+9900 q_1^{10} q_2 q_4+960 q_2^2 q_3 q_4+40 q_4^3+75
q_1^{12} q_5+ 320 q_2^3 q_5+\right.\\ \left.+4 q_3^2 q_5+450 q_1^8
\left(3 q_3 q_4+2 q_2 q_5\right)+140 q_1^6 \left(360 q_2^2 q_4+q_3
q_5\right)+120 q_2 q_4 q_6+ 300 q_1^7 \left(84 q_2^2 q_3+3 q_4^2+4
q_2 q_6\right)+300 q_1^4 \left(16 q_2 q_3 q_4+10 q_2^2
q_5+\right.\right.\\ \left.\left.+q_4 q_6\right)+q_5 q_7+ 10 q_1^3
\left(2720 q_2^3 q_3+420 q_2 q_4^2+q_5^2+180 q_2^2 q_6+q_3
q_7\right)+15 q_1^2 \left(1760 q_2^3 q_4+12 q_3^2 q_4+16 q_2 q_3
q_5+2 q_5 q_6+q_4 q_7\right)+ 60 q_1 \left(320
q_2^5+\right.\right.\\ \left.\left.+2 q_3 q_4^2+q_2 \left(6 q_4
q_5+4 q_3 q_6\right)+q_2^2 \left(12 q_3^2+q_7\right)\right)+30
q_1^5 \left(5600 q_2^4+5 q_4 q_5+q_2 \left(84
q_3^2+q_7\right)\right)\right)$\\

\noindent $\widehat P_{6,8}=12 \left(306000 q_1^{16} q_2^2+25600
q_2^6+9000 q_1^{18} q_3+8 q_3^4+1950 q_1^{12} \left(560 q_2^3+3
q_3^2\right)+ 126000 q_1^{13} q_2 q_4+396000 q_1^9 q_2^2 q_4+960
q_2 q_3 q_4^2+\right.\\ \left.+4950 q_1^{10} \left(80 q_2^2
q_3+q_4^2\right)+7200 q_1^{11} q_2 q_5+8 q_3 q_5^2+ 28800 q_1^7
q_2 \left(3 q_3 q_4+q_2 q_5\right)+120 q_4^2 q_6+70 q_1^6
\left(6720 q_2^3 q_3+24 q_3^3+720 q_2 q_4^2+q_5^2+\right.\right.\\
\left.\left.+480 q_2^2 q_6\right)+ 480 q_2^2 \left(3 q_4 q_5+2 q_3
q_6\right)+900 q_1^8 \left(3080 q_2^4+3 q_4 q_5+3 q_3
q_6\right)+q_7^2+160 q_2^3 \left(12 q_3^2+q_7\right)+ 60 q_1^5
\left(13440 q_2^3 q_4+112 q_2 q_3 q_5+3 q_4 q_7\right)+\right.\\
\left.+40 q_1^3 \left(1920 q_2^2 q_3 q_4+30 q_4^3+800 q_2^3
q_5+240 q_2 q_4 q_6+q_5 q_7\right)+ 150 q_1^4 \left(8960 q_2^5+40
q_2 q_4 q_5+3 \left(12 q_3 q_4^2+q_6^2\right)+4 q_2^2 \left(84
q_3^2+q_7\right)\right)+ \right.\\ \left.+120 q_1^2 \left(1360
q_2^4 q_3+420 q_2^2 q_4^2+120 q_2^3 q_6+q_3 \left(4 q_4 q_5+3 q_3
q_6\right)+2 q_2 \left(q_5^2+q_3 q_7\right)\right)+ 240 q_1
\left(440 q_2^4 q_4+8 q_2^2 q_3 q_5+q_4^2 q_5+q_2
\left(12 q_3^2 q_4+2 q_5 q_6+\right.\right.\right.\\ \left.\left.\left.+q_4 q_7\right)\right)\right)$\\

\noindent $\widehat P_{7,7}= 120 \left(2250 q_1^{23}+34500
q_1^{19} q_2+441600 q_1^{15} q_2^2+3450 q_1^{17} q_3+48300
q_1^{13} q_2 q_3+3450 q_1^{11} \left(520 q_2^3+q_3^2\right)+3450
q_1^{16} q_4+89700 q_1^{12} q_2 q_4+\right.\\ \left.+ 2530
q_1^{10} \left(3 q_3 q_4+2 q_2 q_5\right)+690 q_1^8 \left(720
q_2^2 q_4+q_3 q_5\right)+1150 q_1^9 \left(308 q_2^2 q_3+9 q_4^2+2
q_2 q_6\right)+ 920 q_1^7 \left(3300 q_2^4+21 q_2 q_3^2+2 q_4
q_5+q_3 q_6\right)+\right.\\ \left.+1610 q_1^6 \left(48 q_2 q_3
q_4+16 q_2^2 q_5+q_4 q_6\right)+23 q_1^5 \left(22400 q_2^3 q_3+28
q_3^3+2520 q_2 q_4^2+6 q_5^2+840 q_2^2 q_6+q_3 q_7\right)+115
q_1^4 \left(5600 q_2^3 q_4+16 q_3^2 q_4+\right.\right.\\
\left.\left.+40 q_2 q_3 q_5+2 q_5 q_6+q_4 q_7\right)+ 23 q_1^2
\left(2640 q_2^2 q_3 q_4+60 q_4^3+800 q_2^3 q_5+240 q_2 q_4
q_6+q_5 \left(4 q_3^2+q_7\right)\right)+ 230 q_1^3 \left(4480
q_2^5+14 q_3 q_4^2+q_6^2+\right.\right.\\ \left.\left.+4 q_2
\left(8 q_4 q_5+3 q_3 q_6\right)+2 q_2^2 \left(68
q_3^2+q_7\right)\right)+ 23 q_1 \left(3200 q_2^4 q_3+1200 q_2^2
q_4^2+12 q_3 q_4 q_5+320 q_2^3 q_6+4 q_3^2 q_6+q_6 q_7+4 q_2
\left(4 q_3^3+2 q_5^2+q_3 q_7\right)\right)+\right.\\ \left.+ 92
\left(320 q_2^4 q_4+8 q_2^2 q_3 q_5+q_4 \left(q_4 q_5+q_3
q_6\right)+q_2 \left(8 q_3^2 q_4+2 q_5 q_6+q_4
q_7\right)\right)\right)$\\

\noindent $\widehat P_{7,8}= 390 \left(18000 q_1^{22} q_2+114000
q_1^{18} q_2^2+768000 q_1^{14} q_2^3+20400 q_1^{16} q_2 q_3+19200
q_1^{15} q_2 q_4+4200 q_1^{13} q_3 q_4+1950 q_1^{12}
\left(56 q_2^2 q_3+q_4^2\right)+\right. \\
\left.+240 q_1^{11} \left(780 q_2^2 q_4+q_3 q_5\right)+660
q_1^{10} \left(2600 q_2^4+20 q_2 q_3^2+q_4 q_5\right)+30 q_1^8
\left(12320 q_2^3 q_3+1080 q_2 q_4^2+q_5^2+120 q_2^2 q_6\right)+
200 q_1^9 \left(132 q_2 q_3 q_4+\right.\right.\\ \left.\left.+44
q_2^2 q_5+3 q_4 q_6\right)+160 q_1^7 \left(2880 q_2^3 q_4+9 q_3^2
q_4+12 q_2 q_3 q_5+q_5 q_6\right)+ 560 q_1^6 \left(2640 q_2^5+42
q_2^2 q_3^2+3 q_3 q_4^2+4 q_2 \left(2 q_4 q_5+q_3
q_6\right)\right)+ \right.\\ \left.+8 q_1^5 \left(10080 q_2^2 q_3
q_4+105 q_4^3+2240 q_2^3 q_5+420 q_2 q_4 q_6+q_5 \left(14
q_3^2+q_7\right)\right)+ q_1^2 \left(179200 q_2^6+8 q_3 q_5^2+120
q_4^2 q_6+480 q_2^2 \left(8 q_4 q_5+3 q_3
q_6\right)+\right.\right.\\ \left.\left.+240 q_2 \left(14 q_3
q_4^2+q_6^2\right)+4 q_3^2 q_7+q_7^2+160 q_2^3 \left(68
q_3^2+q_7\right)\right)+ 20 q_1^4 \left(11200 q_2^4 q_3+2520 q_2^2
q_4^2+10 q_3 q_4 q_5+560 q_2^3 q_6+q_6 q_7+2 q_2 \left(28 q_3^3+6
q_5^2+\right.\right.\right.\\ \left.\left.\left.+q_3
q_7\right)\right)+ 80 q_1^3 \left(2800 q_2^4 q_4+40 q_2^2 q_3
q_5+q_4 \left(3 q_4 q_5+4 q_3 q_6\right)+2 q_2 \left(16 q_3^2
q_4+2 q_5 q_6+q_4 q_7\right)\right)+4 \left(1280 q_2^5 q_3+800
q_2^3 q_4^2+160 q_2^4 q_6+\right.\right.\\ \left.\left.+4 q_2^2
\left(4 q_3^3+2 q_5^2+q_3 q_7\right)+q_4 \left(4 q_3^2 q_4+2 q_5
q_6+q_4 q_7\right)+2 q_2 \left(12 q_3 q_4 q_5+4 q_3^2 q_6+q_6
q_7\right)\right)+ 8 q_1 \left(1760 q_2^3 q_3 q_4+4 q_3^3 q_4+400
q_2^4 q_5+q_4 q_5^2+\right.\right.\\ \left.\left.+240 q_2^2 q_4
q_6+q_3 \left(2 q_5 q_6+q_4 q_7\right)+2 q_2 \left(60 q_4^3+q_5
\left(4 q_3^2+q_7\right)\right)\right)\right)$\\

\noindent $\widehat P_{8,8}=15 \left(67500 q_1^{29}+5742000
q_1^{21} q_2^2+77952000 q_1^{13} q_2^4+26100 q_1^{17} \left(760
q_2^3+3 q_3^2\right)+ 4176000 q_1^{14} q_2^2 q_4+1583400 q_1^{12}
q_2 q_3 q_4+\right.\\ \left.+69600 q_1^{15} \left(68 q_2^2 q_3+3
q_4^2\right)+76560 q_1^{10} \left(260 q_2^3 q_4+q_2 q_3
q_5\right)+ 1740 q_1^{11} \left(7280 q_2^3 q_3+13 q_3^3+390 q_2
q_4^2+2 q_5^2\right)+17400 q_1^8 \left(198 q_2^2 q_3 q_4+3
q_4^3+\right.\right.\\ \left.\left.+44 q_2^3 q_5+9 q_2 q_4
q_6\right)+ 32480 q_1^6 q_2 \left(720 q_2^3 q_4+9 q_3^2 q_4+6 q_2
q_3 q_5+q_5 q_6\right)+4350 q_1^9 \left(22880 q_2^5+440 q_2^2
q_3^2+11 q_3 q_4^2+44 q_2 q_4 q_5+2 q_6^2\right)+\right.\\ \left.+
3480 q_1^7 \left(6160 q_2^4 q_3+1080 q_2^2 q_4^2+2 q_2 q_5^2+80
q_2^3 q_6+3 q_3 \left(2 q_4 q_5+q_3 q_6\right)\right)+ 29 q_1^5
\left(1478400 q_2^6+47040 q_2^3 q_3^2+84 q_3^4+10080 q_2 q_3
q_4^2+14 q_3 q_5^2+\right.\right.\\ \left.\left.+630 q_4^2
q_6+6720 q_2^2 \left(2 q_4 q_5+q_3 q_6\right)+3 q_7^2\right)+ 116
q_1^2 \left(33600 q_2^5 q_4+800 q_2^3 q_3 q_5+60 q_2 q_4 \left(3
q_4 q_5+4 q_3 q_6\right)+60 q_2^2 \left(16 q_3^2 q_4+2 q_5 q_6+q_4
q_7\right)+\right.\right.\\ \left.\left.+q_3 \left(30 q_4^3+q_5
q_7\right)\right)+ 290 q_1^3 \left(17920 q_2^5 q_3+6720 q_2^3
q_4^2+36 q_3^2 q_4^2+1120 q_2^4 q_6+6 q_3 q_6^2+8 q_2^2 \left(28
q_3^3+6 q_5^2+q_3 q_7\right)+q_4 \left(4 q_5 q_6+3 q_4 q_7\right)+
\right.\right.\\ \left.\left.+8 q_2 \left(10 q_3 q_4 q_5+q_6
q_7\right)\right)+58 q_1 \left(25600 q_2^7+30 q_4^4+\left(4
q_3^3+q_5^2\right) q_6+160 q_2^3 \left(8 q_4 q_5+3 q_3 q_6\right)+
120 q_2^2 \left(14 q_3 q_4^2+q_6^2\right)+q_4 q_5 \left(8
q_3^2+q_7\right)+\right.\right.\\ \left.\left.+40 q_2^4 \left(68
q_3^2+q_7\right)+q_2 \left(8 q_3 q_5^2+120 q_4^2 q_6+4 q_3^2
q_7+q_7^2\right)\right)+ 290 q_1^4 \left(13440 q_2^3 q_3 q_4+2240
q_2^4 q_5+840 q_2^2 q_4 q_6+q_4 \left(4 q_5^2+3 q_3 q_7\right)+4
q_2 \left(105 q_4^3+\right.\right.\right.\\ \left.\left.\left.+q_5
\left(14 q_3^2+q_7\right)\right)\right)+ 58 \left(1760 q_2^4 q_3
q_4+320 q_2^5 q_5+320 q_2^3 q_4 q_6+q_4 \left(4 q_3 q_4 q_5+q_6
q_7\right)+4 q_2^2 \left(60 q_4^3+q_5 \left(4
q_3^2+q_7\right)\right)+4 q_2 \left(4 q_3^3 q_4+q_4
q_5^2+\right.\right.\right.\\ \left.\left.\left.+q_3 \left(2 q_5
q_6+q_4
q_7\right)\right)\right)\right)$\\
}

\end{document}